\documentclass[12pt]{article}
\usepackage{amsmath}
\usepackage{epsfig}
\usepackage{amsfonts}
\usepackage{amssymb}
\def\beq{\begin{equation}}
\def\eeq{\end{equation}}
\def\bea{\begin{eqnarray}}
\def\eea{\end{eqnarray}}
\def\Tr{{\rm Tr}}
\begin{document}

%----------------------------------------------------------------------
%                     T I T L E
%----------------------------------------------------------------------

\begin{titlepage}

\begin{centering}

\vspace*{3cm}

{\Large\bf On a harmonic solution for 
           two-dimensional adjoint QCD}

\vspace*{1.5cm}

{\bf Uwe Trittmann}
\vspace*{0.5cm}

{\sl Department of Physics\\
Otterbein University\\
Westerville, OH 43081, USA}

\vspace*{1cm}

\today

%----------------------------------------------------------------------%
%                       A B S T R A C T
%----------------------------------------------------------------------%
\vspace*{2cm}

{\large Abstract}

\vspace*{1cm}

\end{centering}

Two-dimensional QCD with adjoint fermions has many attractive
features, yet its single-particle content remains largely unknown. To
lay the foundation for a crucially improved approximation of the
theory's spectrum, we developed a method to find the basis of
eigenstates using the symmetry structure of the asymptotic theory
where pair production is disallowed.  This method produces complete
sets of multi-dimensional harmonic functions for the massless and the
massive theory. Previously only part of such a basis was known.  The
method presented here should be applicable to other theories
and has the promise of factoring out the long-range Coulomb-type part of
interactions.  The role of pair production and implications for the
bosonized theory in the case of adjoint QCD$_2$ are discussed.
\vspace{0.5cm}

%\noindent
%PACS number(s): 

\end{titlepage}
\newpage

%----------------------------------------------------------------------%
%                       Introduction
%----------------------------------------------------------------------%

\section{Introduction}

Two-dimensional adjoint quantum chromodynamics, QCD$_{2A}$, is 
a non-abelian Yang-Mills theory 
coupled to fermions in the adjoint representation, and
based on the Lagrangian
\beq\label{Lagrangian}
{\cal L}=Tr[-\frac{1}{4g^2}F_{\mu\nu}F^{\mu\nu}+
i\bar{\Psi}\gamma_{\mu}D^{\mu}\Psi],
\eeq
where $\Psi=2^{-1/4}({\psi \atop \chi})$, 
with $\psi$ and $\chi$ 
being $N\times N$ matrices. The field strength is
$F_{\mu\nu}=\partial_{\mu}A_{\nu}-\partial_{\nu}A_{\mu}+i[A_{\mu},A_{\nu}]$,
and the covariant derivative is defined as $D_{\mu}=\partial_{\mu}
+i[A_{\mu},\cdot]$.
Throughout the paper,
light-cone coordinates $x^\pm=(x^0\pm x^1)/\sqrt{2}$ are used,
where $x^+$ plays the role of a time. We will work in the light-cone gauge,
$A^+=0$, effectively omitting fermionic zero modes. 
The theory is discussed extensively in the literature, so we refer the
reader to Refs.~\cite{Kutasov94, DalleyKlebanov, BDK, GHK, Katz} for
details. Two main features are an asymptotic supersymmetry, and
the stark contrast between the massive and the massless versions of the
theory \cite{LesHouches}. Recently, there has been interest in three- and
four-dimensional versions of the theory \cite{AdjQCD3,AdjQCD4}.

While the ultimate goal is a solution of the full theory, our
aim here is much more modest. As a foundation for future work, we
construct in Sec.~\ref{SecBasis} a
complete set of eigenstates
for the {\em asymptotic} theory based on symmetries which are broken
in the full theory, thereby expanding work begun in \cite{UT3}.
The basis is a set of linear
combinations of multi-dimensional harmonic functions subject to
the symmetry constraints furnished by the light-cone Hamiltonian
derived from Eq.~(\ref{Lagrangian}).  Since the asymptotic theory has
disjoint sectors with different parton numbers, the basis states can
be organized by parton number $r$, which results in $r-1$ excitation numbers
responsible for the multi-dimensionality of the solution.
Technically, the basis will be appropriate only in the high excitation number
limit. Empirically, we find that even the lowest states are well
represented, probably due to their large separation in (bound state)
mass.  In the full theory
particle pair production couples the disjoint parton sectors,
and eigenfunctions of the full theory will be linear combinations of
basis states of different parton numbers. Thus, the harmonic basis
states can be used to discuss the role of particle creation and
annihilation, as envisioned in \cite{Kutasov94}. This is done
in Sec.~\ref{SecPairProduction}.
%With parton sector mixing in place,
%we can study the emergence of multi-particle states. In a fermionic
%theory, their existence is guaranteed because because the system can
%be bosonized, which requires a fine-tuned symmetrization of fermionic
%into bosonic states. We investigate this phenomenon and the
%implications for finite discretizations of the theory in
%Sec.~\ref{SecMPS}.
As applications of our method we 
briefly describe a program to solve the full theory with 
numerical methods in Sec.~\ref{SecUsing_eLCQ}, and point out implications
for the bosonized version of the theory in Sec.~\ref{SecBosonized}.
Finally, we discuss the results and general applicability of the method
in Sec.~\ref{SecDiscussion}.

%------------------------------------------------------------
\section{Constructing a Harmonic Basis}
\label{SecBasis}
%------------------------------------------------------------

\subsection{Introductory Remarks}

Starting from the QCD$_{2A}$ Lagrangian, Eq.~(\ref{Lagrangian}), the
dynamics of a system of adjoint fermions interacting via
a non-dynamical gluon field in two dimensions can be described by a
light-cone momentum operator $P^+$
and Hamiltonian operator $P^-$. The two operators are expressed
in terms of fermionic operators subject to the
anti-commutation relation
\beq\label{Commy}
\{b_{ij}(k^{+}), b_{lk}^{\dagger}(p^{+})\} =
\delta(k^{+} - {p}^{+})
(\delta_{il} \delta_{jk}-\frac{1}{N}\delta_{ij} \delta_{kl})\,\,.
\eeq
To wit
\begin{eqnarray}
P^+ &=& \int_{0}^{\infty} dk\ k\, b_{ij}^{\dagger}(k)b_{ij}(k)\ ,\\
P^{-} &=& {m^2\over 2}\, \int_{0}^{\infty}\label{P_BDK}
{dk\over k} b_{ij}^{\dagger}(k)
b_{ij}(k) +{g^2 N\over \pi} \int_{0}^{\infty} {dk\over k}\
C(k) b_{ij}^{\dagger}(k)b_{ij}(k) \\
&&+ {g^2\over 2\pi} \int_{0}^{\infty} dk_{1} dk_{2} dk_{3} dk_{4}
\biggl\{ B(k_i) \delta(k_{1} + k_{2} +k_{3} -k_{4})\nonumber \\
&&\qquad\qquad\times(b_{kj}^{\dagger}(k_{4})b_{kl}(k_{1})b_{li}(k_{2})
b_{ij}(k_{3})-
b_{kj}^{\dagger}(k_{1})b_{jl}^{\dagger}(k_{2})
b_{li}^{\dagger}(k_{3})b_{ki}(k_{4})) \nonumber\\
&&\qquad + A(k_i) \delta (k_{1}+k_{2}-k_{3}-k_{4})
b_{kj}^{\dagger}(k_{3})b_{ji}^{\dagger}(k_{4})b_{kl}(k_{1})b_{li}(k_{2})
 \nonumber\\
&&\qquad + \frac{1}{2} D(k_i) \delta (k_{1}+k_{2}-k_{3}-k_{4})
b_{ij}^{\dagger}(k_{3})b_{kl}^{\dagger}(k_{4})b_{il}(k_{1})b_{kj}(k_{2})
\biggl\}, \nonumber
\end{eqnarray}
with
\begin{eqnarray}
A(k_i)&=& {1\over (k_{4}-k_{2})^2 } -
{1\over (k_{1}+k_{2})^2}\ , \label {A} \label{EqnA}\\
B(k_i)&=& {1\over (k_{2}+k_{3})^2 } - {1\over (k_{1}+k_{2})^2 }, \label {B}\\
C(k)&=& \int_{0}^{k} dp \,\,{k\over (p-k)^2},\\ 
D(k_i)&=& \frac{1}{(k_{1}-k_{4})^2} - \frac{1}{(k_{2}-k_{4})^2},
\end{eqnarray}
where the trace-splitting term $D(k_i)$ can be omitted at large $N_c$, and 
the trace-joining term is proportional to $B(k_i)$. 
The structure of the Hamiltonian $P^-$ displayed in Eq.~(\ref{P_BDK}) is 
\beq\label{StructureOfHamiltonian}
P^-= P^-_{m}+P^-_{ren}+ P^-_{PC,s}+P^-_{PC,ns}+P^-_{PV} + P^-_{finiteN} .
\eeq
The mass term $P^-_{m}$ is dropped in the massless theory, but the 
renormalization operator $P^-_{ren}$ needs to be included. 
Parton-number violating terms, $P^-_{PV}$, couple blocks of 
different parton number. Parton-number conserving interactions 
$P^-_{PC}$ are block diagonal, and may include singular($s$) or
non-singular($ns$) functions of the parton momenta.
For details see \cite{DalleyKlebanov,BDK,UT3}.

The problem is cast into an eigenvalue equation
\beq\label{EVP}
2P^+P^-|\Psi\rangle\equiv H_{LC}|\Psi\rangle=M^2|\Phi\rangle.
\eeq
Namely, the light-cone Hamiltonian $H_{LC}$ acts on an eigenket $|\Psi\rangle$
yielding the mass (squared) of a bound state as the eigenvalue.
The eigenkets are in general linear combinations of states of definite
parton(fermion) number $r$
\beq\label{TheStates}
|\Phi_r\rangle=
\int_0^{\frac{1}{r}} dx_1
\left(
\prod^{r-1}_{i=2}
\int^{1-(r-1)x_1-\sum^{i-1}_{j=2}x_j}_{x_1}dx_i
\right)
%\left(\prod_{j=1}^r\int_0^1 dx_j\right)\,
%\delta\left(1-\sum^r_{i=1} x_i\right)
\frac{\phi_r(x_1,x_2,\ldots,x_r)}{N_c^{r/2}}
Tr[b(-x_1)\cdots b(-x_r)]|0\rangle.
\eeq
The wavefunctions $\phi_r$ distribute momentum between the partons.
Note that there are only $r-1$ integrations since total momentum can be
set to unity. The integration is over the effective Hilbert space, which
looks complicated due to elimination of redundant operator combinations like
$\Tr[b(x)b(y)b(z)]=\Tr[b(y)b(z)b(x)]$. The
explicit shape of the integration domain is displayed to emphasize that
for $r>3$ the wavefunction cannot be reconstructed\footnote{Meaning we cannot
  determine the wavefunction in this region from symmetries
  and its values close to the domain boundaries. For $r=2$ we can:
  knowing $\phi_2(x)$ in $[0,\frac{1}{2}]$ and $\phi_2(x)=-\phi(1-x)$
is obviously enough.} in the
region around the middle of the naive Hilbert space $[0,1]^{r-1}$.

In the asymptotic limit, where only highly excited states are considered,
both parton-violation and mass terms can be neglected.
As a consequence the asymptotic theory splits into decoupled sectors with 
fixed parton numbers subject to 't Hooft-like equations
\beq
\frac{M^2}{g^2N}\phi_r(x_1,\ldots,x_r)=-\sum_{i=1}^r(-1)^{(r+1)(i+1)}
\int_{-\infty}^{\infty}\frac{\phi_r(y,x_i+x_{i+1}-y,
x_{i+2},\ldots,x_{i+r-1})}{(x_i-y)^2}dy.
\label{TheEquation}
\eeq
The total 
momentum is set to unity, and thus the momentum fractions $x_i$ 
add up to one, $\sum_i x_i=1$. Clearly,
the number of partons $r$ is even (odd) for bosonic (fermionic) states. 
In \cite{Kutasov94,UT3}, additionally the approximation
\beq\label{approx}
\int^1_0\frac{dy}{(x-y)^2}\phi(y) \approx 
\int^\infty_{-\infty}\frac{dy}{(x-y)^2}\phi(y) 
\eeq
was used. Mathematically this is helpful because the solutions
of the eigenvalue problem then are harmonic functions, see
Eq.~(\ref{r2tHooft}). The approximation makes sense physically, 
because for the highly excited states the integral is dominated by the 
interval around $x=y$ which is associated with the long-range
Coulomb-type force.  

In \cite{UT3} we showed that the integral equation (\ref{TheEquation})
can be solved algebraically using the ansatz 
\beq\label{Ansatz}
|n_1, n_2,\ldots n_{r-1} \rangle \doteq \prod^{r-1}_j e^{i \pi n_j x_j}
=\phi_r(x_1,x_2,\ldots,x_r),
\eeq
where $x_r=1-\sum_j^{r-1} x_j$. 
The ansatz is motivated by its simplest ($r=2$) version, which
solves the 't Hooft equation of {\em fundamental} QCD$_2$ \cite{tHooft} 
\beq\label{r2tHooft}
\frac{M^2}{g^2 N} e^{i\pi n x} = -\int^{\infty}_{-\infty} 
\frac{dy}{(x-y)^2} e^{i\pi n y} = \pi |n| e^{i\pi n x}, 
\eeq
where the excitation number $n$ is integer.
We thus use the single-particle states of a Hamiltonian
appropriate for the problem to construct a Fock basis,  
inspired by \cite{Pauli84}.  

Since the integral equation (\ref{TheEquation}) is more involved than the
't Hooft equation, we have to symmetrize the ansatz (\ref{Ansatz}) to comply
with the constraints inherent in the Hamiltonian (\ref{EVP}).
Namely, the solutions of the adjoint 't Hooft problem have to be
(pseudo-)cyclic,
\beq\label{Cyclicity}
\phi_r(x_1,x_2,\ldots,x_r)=(-1)^{r+1}\phi_r(x_2,x_3\ldots,x_r,x_1),
\eeq
since the fermions are real. To implement this constraint we introduce the
cyclic permutation operator
\[
{\cal C}: (x_1,x_2,\ldots,x_r)\rightarrow(x_2,x_3,\ldots,x_r,x_1).
\]
Since the Hamiltonian is unchanged by a color index reversal of its operators,
the string or trace of fermionic operators in the states (\ref{TheStates})
can be reversed at will.
The solutions can therefore be organized into sectors of definite parity under
the orientation symmetry,
\beq\label{Tsymmetry}
{\cal T}: b_{ij} \rightarrow b_{ji}.
\eeq
The two symmetry operators act on the ansatz, Eq.~(\ref{Ansatz}), as follows.
\bea\label{CTSymmetryAction}
{\cal C}&:& |n_1,n_2,\ldots,n_{r-1}\rangle 
\rightarrow (-1)^{n_{r-1}} |-n_{r-1},n_1-n_{r-1},n_2-n_{r-1},\ldots, n_{r-2}-n_{r-1}
\rangle, \nonumber
\\
{\cal T}&:& |n_1,n_2,\ldots,n_{r-1}\rangle 
\rightarrow (-1)^{n_1} |-n_1,n_{r-1}-n_1,n_{r-2}-n_1,\ldots, n_2-n_1 \rangle.
\label{CTaction}
\eea
In \cite{UT3} we constructed eigenfunctions
in the two- and three-parton sectors ($r=2,3$) for both the massive and
the massless theory, with bound-state masses
\beq\label{AnsatzMasses}
M^2 = g^2 N\pi \left(|n_1|+|n_{r-1}|+\sum_{k=1}^{r-2}|n_k-n_{k+1}|\right).
\eeq
The crucial ingredient of the method is the symmetrization due to ${\cal C}$,
\beq\label{CsymAnsatz}
|n_1,n_2, \ldots n_{r-1}\rangle_{sym}\equiv \frac{1}{\sqrt{r}}
\sum_{k=1}^r (-1)^{(r-1)(k-1)}{\cal C}^{k-1}|n_1,n_2, \ldots n_{r-1}\rangle,
\eeq
where ${\cal C}^{0}=1$, because only the symmetrized states reflect
the (pseudo)cyclic structure of the Hamiltonian.

While the $r=2,3$ wavefunctions reproduce known (DLCQ) solutions of the
theory remarkably well down to the lowest states, the ansatz
fails to work for $r>3$ in general. In \cite{UT3}, we were able to 
express a subset of the $r=4$ solutions as a linear combination of the
symmetrized four-parton states. These solutions coincide with the
ones derived earlier \cite{Kutasov94}. To introduce notation and
show the complexity of the problem, we display a
five-parton state symmetrized under ${\cal C}$ and ${\cal T}$, i.e. of
definite $C$ and $T$ quantum numbers\footnote{While $T$ is an ordinary quantum
  number reflecting a symmetry of the Hamiltonian, $C$ (or rather the set of
  $C_i$) is fixed by the
  constraint that the state has an (anti-)cyclic wavefunction,
  Eq.~(\ref{Cyclicity}), as required by the structure of the Hamiltonian as
  a sum over permutations of the parton momenta.}.
Note that the state consists of $4r=20$ {\em statelets},
characterized by an $(r-1)$-tuple
of (ordered) excitation numbers $n_i$ (here: $(n,m,l,k)$)
{%\small
\bea\label{phi5}
&& \!\!\!\!\!\!\!\!\!\!|\phi_5,n,m,l,k; 
\bar{M}^2=|n|+|n-m|+|m-l|+|l-k|+|k|\rangle_{T}=\\
&&
|n,m,l,k\rangle +(-1)^k |-k,n-k,m-k,l-k\rangle+(-1)^l|k-l,-l,n-l,m-l\rangle
\nonumber\\
&&\quad +(-1)^m |l-m,k-m,-m,n-m\rangle+(-1)^n |m-n,l-n,k-n,-n\rangle
\nonumber\\  
&&\!\!\!\!\!\!\!+T\left[(-1)^n |-n,k-n,l-n,m-n\rangle 
  + |k,l,m,n\rangle+(-1)^k |l-k,m-k,n-k,-k\rangle\right.\nonumber\\
&&\quad \left. +(-1)^l |m-l,n-l,-l,k-l\rangle+(-1)^m |n-m,-m,k-m,l-m\rangle  
\right].\nonumber
\eea 
}
Typically, these states are paired with a partner state of negative excitation
numbers to create a real wavefunction, i.e.~a sine or cosine.
It is convenient to do so with an additional symmetry
in mind, which we'll introduce in the next section.

\subsection{
  %A new Interpretation of an old Constraint Leads to
  Exhaustive Symmetrization}
\label{SecExhaustive}

%%%%%%%%%%%%%%%%%%%%%%%% NEW WORK
%One ingredient necessary to finish the construction of a
%complete basis apparently has been overlooked thus far --- or at least
%glossed over. 
As is well known, the eigenfunctions of the massive theory have
to vanish when one parton momentum is zero to guarantee hermiticity of
the Hamiltonian \cite{tHooft}
\beq\label{ZeroBC}
\phi_n(0,x_2, \ldots, x_n)=0.
\eeq
This is sometimes called a boundary
condition because it behaves as one, but this is
misleading --- after all, we are trying to solve an integral and not
  a differential equation. 
In the massive two-parton sector (or the simpler fundamental theory $QCD_{2f}$
\cite{tHooft}) this leads to sine eigenfunctions,
and to cosine eigenfunctions in the massless theory. Clearly, the
general verdict is that the Hilbert space of the theory
splits into two disjoint sets of functions: one even and the other one odd.
It seems that the consequences of this straightforward observation have not
been fully realized.
This is not surprising, since the answer
to the simple question --- under which transformation or symmetry
the eigenfunctions are odd or even --- is trivial for few partons, and
fairly complicated for many. 
Indeed, the symmetry in question is manifest only in the parton sectors with
$r\ge 4$, because it is redundant with ${\cal C}$ and ${\cal T}$ otherwise.
To get a handle on it, note that while
{physically} it is true that we need to manage the behavior of
the wavefunctions at the boundaries of the domain of integration,
{mathematically} {\em boundary conditions} are {not} the right
tool.
%--- simply because we cannot specify boundary conditions in an integral
% equations.
%
Rather, we have to implement {\em symmetries} that will result in the
desired behavior of the wavefunctions at the boundary 
%\footnote{
%  Of course, we cannot ``implement'' symmetries.  Rather, we have to
%  make sure that the state has the correct behavior at the
%  ``boundary'' which is guaranteed by its properties under some symmetry
%  ${\cal S}_i$
  --- just like the ${\cal C}$ symmetrization assures that the
  state so constructed is an eigenstate of the Hamiltonian.
% }.
We alluded to such a symmetry in \cite{UT3}, % Eqn. (19)
but the general method is more involved.

For massive fermions, we need the wavefunctions to vanish on the hyperplanes
characterized by $x_i=0$ for at least one $x_i$.
%Hence, for every term in our modular
Since our wavefunction ansatz (\ref{Ansatz}) is modular, we can simply add
for every term of the form $e^{i\pi\sum_j n_jx_j}$
another term with opposite sign that is the same for $x_i=0$
but different\footnote{Because the wavefunction would otherwise
be identically zero, of course.} for $x_i\neq 0$. 
This idea can be realized by introducing $r-1$ symmetry operators
${\cal S}_i$ which we might call lower-dimensional
inversions,
%\footnote{We use $\cal S$, for the German word ``Spiegelung''.}
because they invert all but one of the excitation numbers
{\small
\beq\label{LDI}
  {\cal S}_i:
  |n_1,n_2,\ldots, n_i\ldots,n_{r-1}\rangle
  \rightarrow|-n_1,-n_2,\ldots,n_i-n_{i+1}-n_{i-1}(1-\delta_{1i}),
  \ldots,-n_{r-1}\rangle.
  \eeq
  }
The replacement of the $i$th excitation number is such that the
mass (squared) of the
state remains invariant, see Eq.~(\ref{AnsatzMasses}).
%With string reversal ${\cal T}$ in place, it is only necessary to
%symmetrize half of the excitation numbers with ${\cal S}$. Hence
This symmetry is hidden in the
three-parton sector $r=3$ because the low-dimensional inversion
can be expressed in terms of the other symmetry operations\footnote{At $r=2$
  the situation is fully degenerate with ${\cal C}={\cal T}={\cal S}$ up to
  signs.}
(${\cal S} = {\cal TC}^2$, see Eq.~(26) of \cite{UT3}), so that $r=4$ is
the lowest parton sector in which the full symmetry unfolds\footnote{This
  has to be taken with a grain of salt, since at $r=4$ we have
  ${\cal S}_1{\cal T}{\cal C}^2={\cal S}_2{\cal S}_1$, so there are only
  two independent ${\cal S}$ operators, not $N(r)=\frac{1}{2}(r-1)!-1$, see
  below.}.

In Sec.~\ref{SecGroupTheory} we will see that additional ${\cal S}$
operators creep in, so
let's call the set of lower-dimensional inversions
${\cal E}=\{{\cal S}_1, {\cal S}_2, \ldots, {\cal S}_{N(r)}\}$.
For a given parton number $r$,
the maximal number $N(r)$ of independent ${\cal S}$ operators can be
determined from a group theoretical argument. 
%where $\bar{r}=r-1$ for odd $r$
%and simply $r$ otherwise.
In practice, the argument boils down to combinatorics:
the number of operators of the permutation group of order $r$ with inversion
is $2 r!$ We derive the full symmetry group
%\footnote{${\cal G}$ might stand for {\em Gesamtsymmetrisierungsgruppe}.}
$\cal G$ in
Sec.~\ref{SecGroupTheory}, and use only its generic properties here.
The group choice is intuitive as we will see, since in the $r$ parton
sector, we are in essence permuting $r$ objects. That the objects (the partons)
carry different momenta is irrelevant here. We are thus symmetrizing
the wavefunctions under the set of $N(r)$ ${\cal S}$-symmetries. Later
we will see that there is only one common multiplicative
$Z_2$ quantum number
%\footnote{That is, $S^2=1$.}
$S$.
Tentatively, we write
\[
|r\rangle_{Ssym} = \big(1+ \sum_i^{N(r)}
S_i{\cal S}_{i=1}\big)|r \rangle_{sym}, 
\]
where $|r\rangle_{sym}$ is the ansatz state $|n_1,n_2,\ldots,n_{r-1}\rangle$
symmetrized under the cyclic group $\langle{\cal C}\rangle$,
Eq.~(\ref{CsymAnsatz}).

To include all symmetries in our approach, it is convenient to make the
inversion of excitation numbers explicit with the operator
\beq\label{Inverse}
  {\cal I}:
  |n_1,n_2,\ldots,n_{r-1}\rangle\rightarrow|-n_1,-n_2,\ldots,-n_{r-1}\rangle.
\eeq
Clearly, ${\cal I}$ is a $Z_2$ operator,
and states even and odd under ${\cal I}$
simply represent cosine and sine wavefunctions, respectively.
We can then write down an orthonormal set of basis states in all sectors
of the theory, characterized by their $Z_2$ quantum numbers $(T,I,S)$ under
the symmetry transformations ${\cal T}$, ${\cal I}$, and ${\cal S}$, and their excitation numbers $n_i$ collected in $|r\rangle$
\beq\label{TheSolution}
|r\rangle_{FullSym} \equiv {\cal G}|r\rangle= \left\{\frac{\left(1+T{\cal T}\right)\left(1+I{\cal I}\right)}
            {\sqrt{2r!{\cal N}}}
            % r from C, 2 from T, another 2 from I, 1+bar{r}/2 from S
            % NO! r from C, 2 from T, another 2 from I, 2^{bar{r}/2} from S
\sum_{k=0}^{r-1} (-)^{(r-1)k}{\cal C}^k
\big(1+\sum_i^{N(r)} S_i{\cal S}_{i=1}\big)|r \rangle
\right\},
\eeq
where ${\cal N}$ is the volume of the Hilbert space in the $r$ parton sector,
\beq\label{HSVolume}
{\cal N}=\int_0^{\frac{1}{r}} dx_1\left(\prod^{r-1}_{i=2}
\int^{1-(r-1)x_1-\sum^{i-1}_{j=2}x_j}_{x_1}dx_i\right)=\frac{1}{r!},
\eeq
{\em cf.}~Appendix of Ref.~\cite{UT3}.
The excitation numbers can, in general, be even or odd integers.
Since the states, Eq.~(\ref{TheSolution}),
are by construction eigenstates of the light-cone Hamiltonian
with masses given by Eq.~(\ref{AnsatzMasses}), they furnish, in principle,
a full solution\footnote{This construction should also
  settle the issue raised by 't Hooft in the footnote of his seminal paper
  \cite{tHooft}: since the ``boundary conditions'' are
  really %(space-time)
  symmetrizations, they must hold order by order in $N_c$.}
% DANGER
of asymptotic QCD$_{2A}$ --- keeping in mind that we made the
approximation, Eq.~(\ref{approx}).

\subsection{Classification of States}

In practice, most sectors prescribed by Eq.~(\ref{TheSolution}) are empty.
As can be gleaned from the symmetry operations, Eqs.~(\ref{CTSymmetryAction}),
(\ref{LDI}), and (\ref{Inverse}), only two sets of quantum
numbers lead to viable states. Namely, only states with all even
excitation numbers, and states with alternating odd and even
numbers give rise to {\em bona fide} states. In the latter case, the first and
last excitation numbers must be odd, so that these states only exist
in the even parton sectors. Even so, most of these excitation number
combinations are not viable due to cancellations of terms. For example,
{\em bona fide} states are only found in the $(T\pm,I\mp)$ four-parton sectors
and in the $(T\pm,I\pm)$ five-parton sectors, see Table \ref{StateTable}. In
the odd parton sectors, these four sectors (for both
$S\pm$) are indeed the four sectors necessary to describe the two
${\cal T}$-sectors of the massive and the massless theory, respectively.
For an even number
of partons, the addition of the mixed excitation number states
($|{odd, even, odd}\rangle\equiv|oeo\rangle$ for $r=4$) does,
of course, not lead to more sectors.
Rather, two even excitation number sectors do not yield
{\em bona fide} states. For an explanation, see Appx.~\ref{DerivationOfGMT}.
For instance, in the four parton sector, only
$(T\pm,I\mp,S\pm)$ in the even excitation number sector and $(T\pm,I\mp,S\mp)$ with mixed
even and odd $n_i$ are viable; the former are
the two sectors of the massive theory, and the latter represent
the two massless sectors.
This is a straightforward generalization from the
earlier findings \cite{UT3} that in the two-parton sectors cosines
with odd excitation number represent the massless and sines with even
excitation number the massive theory, whereas all excitation numbers
in the three-parton sector are even.

%%%%%%%%%%%%%%%%%%%% Massless Four %%%%%%%%%%%%%%%%%
%
\begin{figure}
\centerline{
\psfig{file=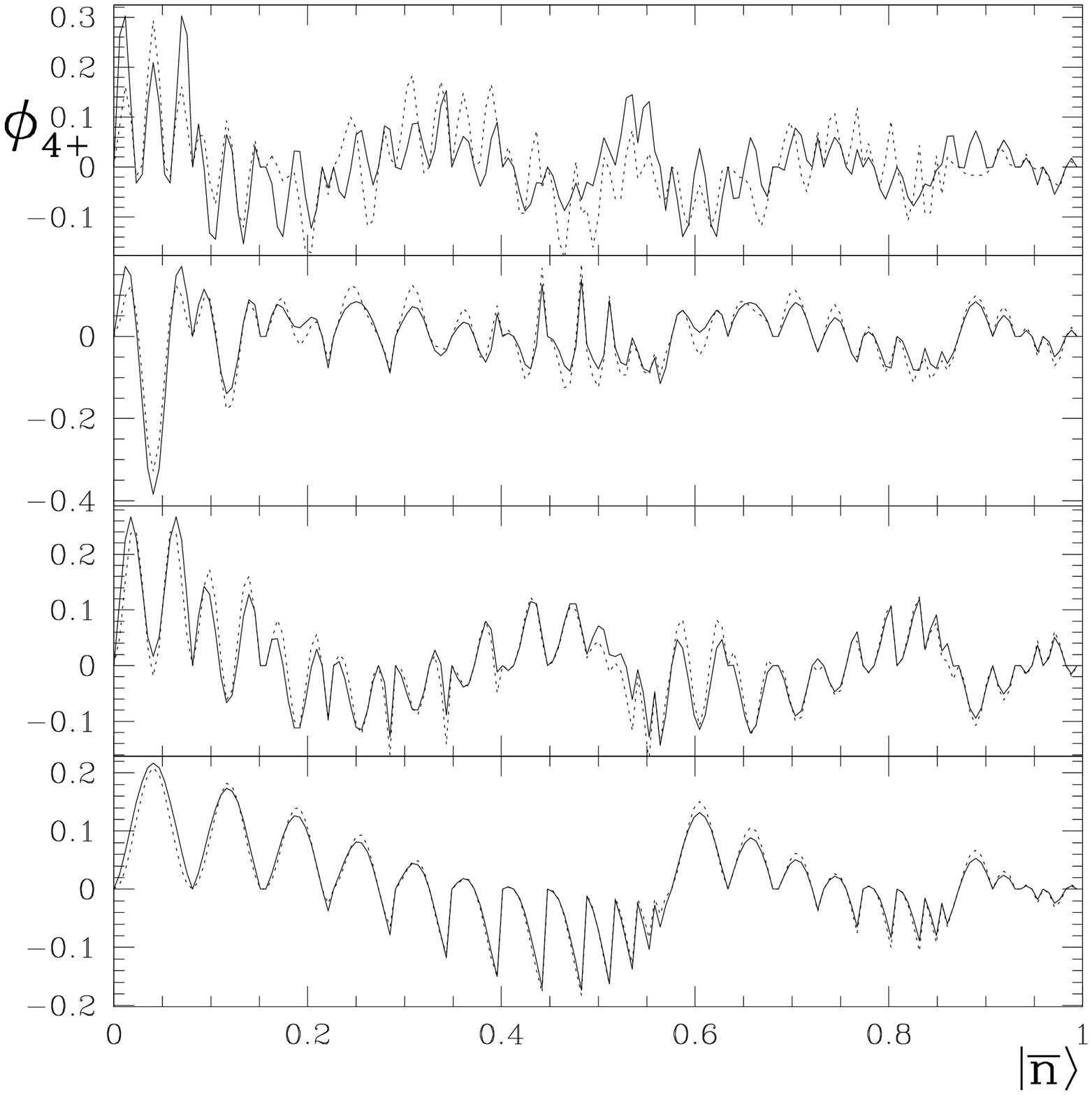,width=7.8cm}
\psfig{file=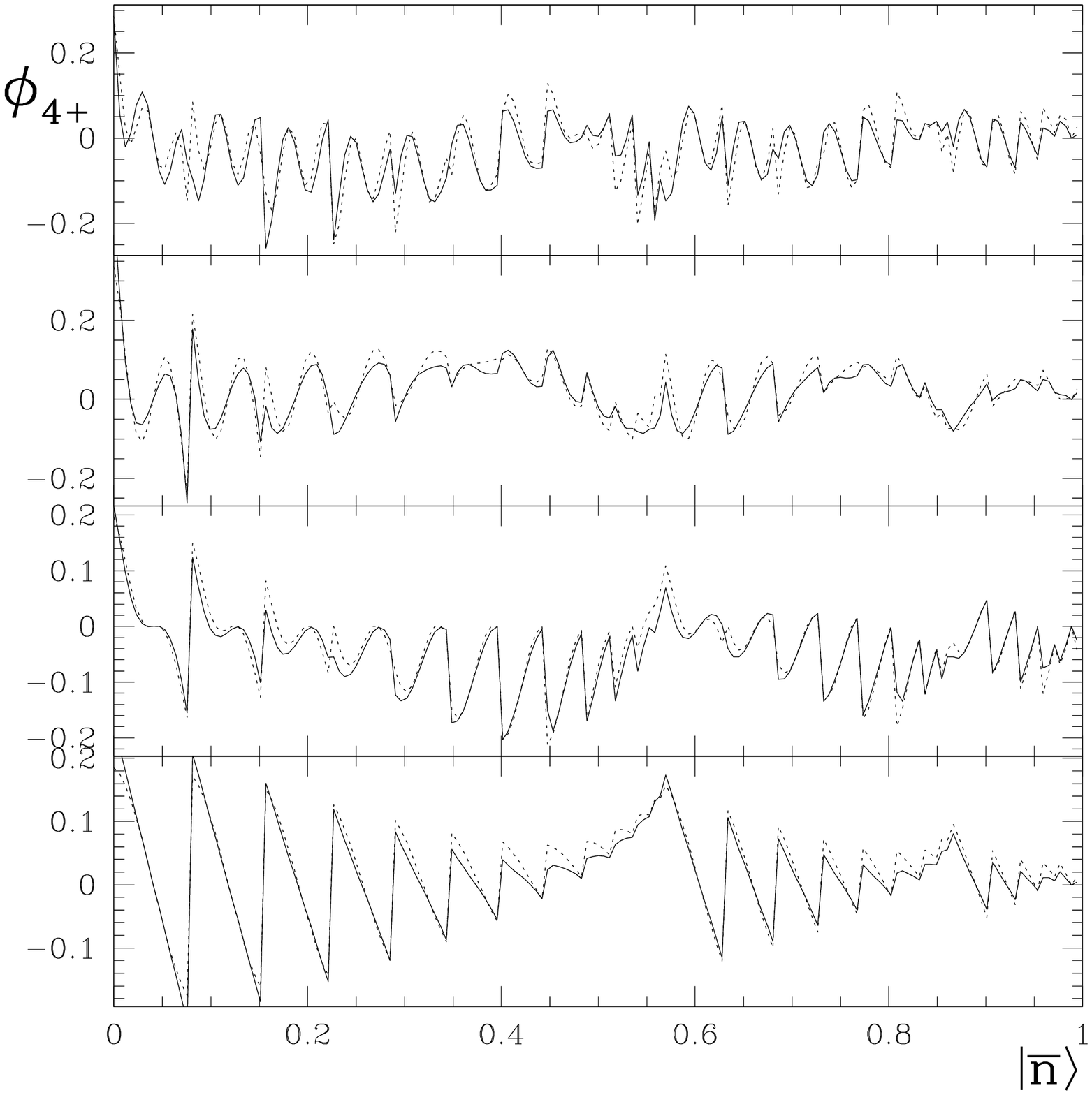,width=7.8cm}
}
\caption{Lowest DLCQ (solid lines, $K=32$) and asymptotic 
  eigenfunctions (dashed lines) in the {\em massless} four-parton sectors,
  i.e.~$|oeo\rangle$ states with $\mu\equiv m^2/g^2N=0$,
  plotted as a function of the normalized
  basis state number $\bar n$. 
  Left (a):  ${ T}+$, hence ${ I}-{ S}-$.
  Right (b):  ${ T}-$, hence ${ I}+{ S}+$.
  We used a rather low harmonic resolution $K$ to increase
  readability,
  since we do not need to worry about convergence here.
\label{Fig4partons}}
\end{figure}
%
%%%%%%%%%%%%%%%%%%%%%%%%%%%%%%%%%%%%%

To check our musings, we make contact with known results. Our solution
(\ref{TheSolution}) reproduces the results of Ref.~\cite{Kutasov94}
Eq.~(4.13) and classifies them as $|eee; T-I+S-\rangle$ states.  The
wavefunctions (4.13) of \cite{Kutasov94} exhibit only two (not
three) excitation numbers, because they represent a subset of the full
set of wavefunctions.
%(Which ones?)
We thus find that there are more states
than anticipated, and that the counting of states is more
involved; it does not seem to lend itself to a string-motivated
parametrization.
%(MORE, in particular figure out an algorithm to
%easily generate the basis states!)
Note that the solution (4.13) in
\cite{Kutasov94} is much more compact and looks
differently (double sines versus triple cosines), but this is a superficial
disagreement and the price one
has to pay for generality: 12 terms\footnote{Obviously, a sine has {\em two}
  exponential terms.}  of (4.13) in \cite{Kutasov94} vs.~48
terms in Eq.~(\ref{TheSolution}) at $r=4$.
%One might worry
%that this pattern renders the approach
%  unusable due to the factorial rise of the number of
%  terms. Fortunately, $r=4$ already exhibits the full extend of
%  symmetrization needed\footnote{But see footnote above.  As a
%    consequence, there are ``untapped'' symmetries in the higher
%    parton sectors. In principle, their wavefunctions could be
%    classified by these symmetries, hence the Hamiltonian
%    block-diagonalized. We thus conclude that the states in different
%    higher symmetry sectors do not interact --- according to the
%    Wigner-Eckard theorem.}, and instead of $2\times r!$, the number
%  of terms only grows like
%% NO! $2(2+\bar{r})r=48,60,96,112,160,\ldots$
%$4r 2^{\bar{r}/2}=80,196,224,512,\ldots$
%for $r>4$, which is, of course, still an exponential growth. 
Next we check how accurate our algebraic solution
is by comparing to a numerical (DLCQ) calculation, see
Fig.~\ref{Fig4partons}. Some masses are degenerate at $r=4$ as opposed to
$r<4$, and the agreement is not
as good as in the three-parton sector \cite{UT3}, due to considerable
mixing of states of equal mass. Recall that the algebraic
eigensolutions were derived by making the approximation
Eq.~(\ref{approx}); using the correct limits of the integral
evidently induces interactions between the algebraic basis states.
As can be gleaned from Fig.~\ref{Fig4partons}, the discrepancy
between algebraic and numerical solutions is noticeable even
in the massless sector, where the masses are {\em not} degenerate.
This heralds the worsening of our approximation at large parton number $r$.
Namely, 
the effective volume of the Hilbert space decreases as $1/r!$, see  
Eq.~(\ref{HSVolume}). As a consequence, the integral looks less and less
like $(\int^{\infty}_{-\infty}dx)^{r-1}$, although the core motivation for 
the approximation (that the region around the singularity is
the most important) remains valid.

%%%%%%%%%%%%%%%%%%%% Massive Four %%%%%%%%%%%%%%%%%
%
\begin{figure}
\centerline{
\psfig{file=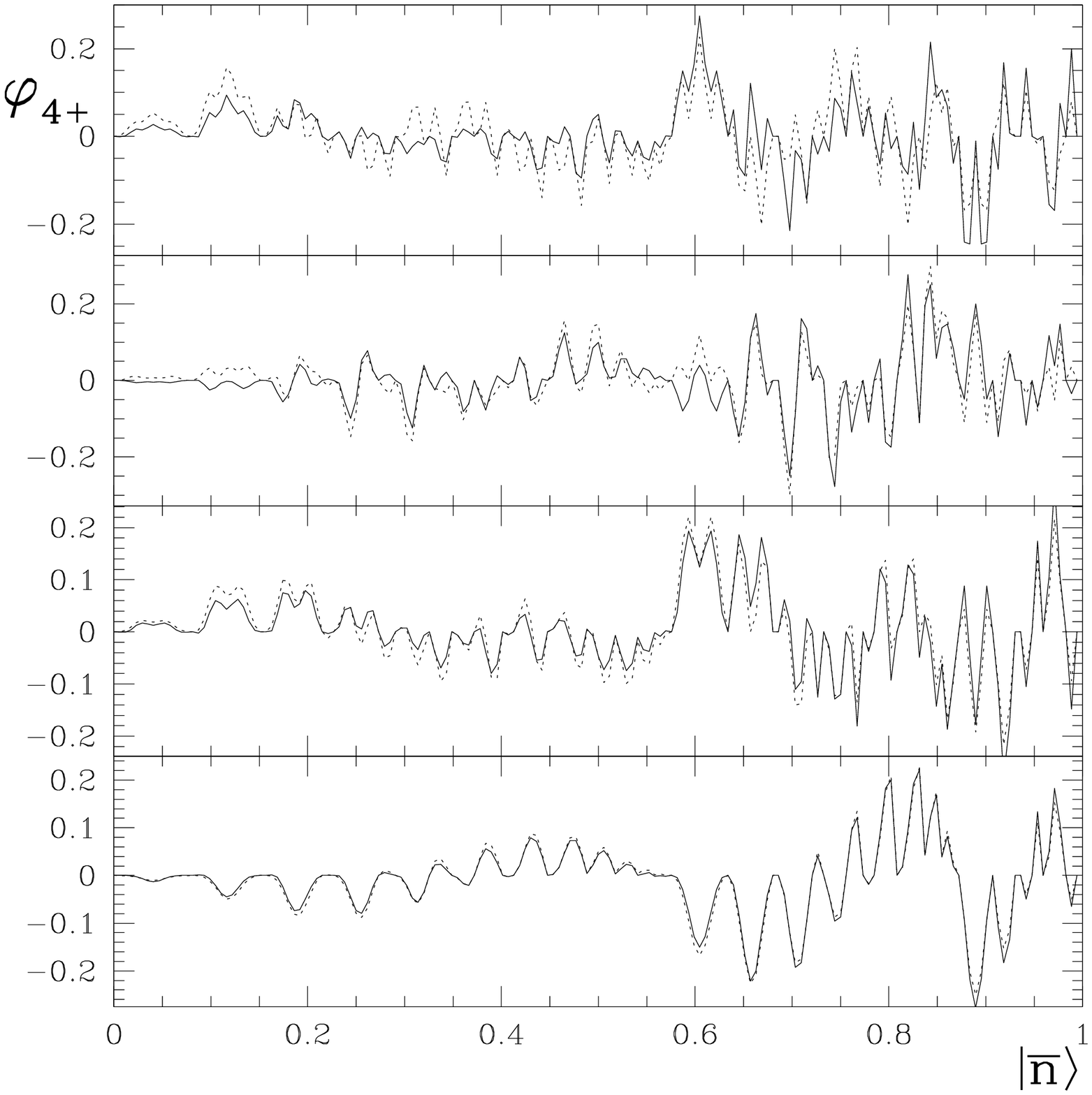,width=7.8cm}
\psfig{file=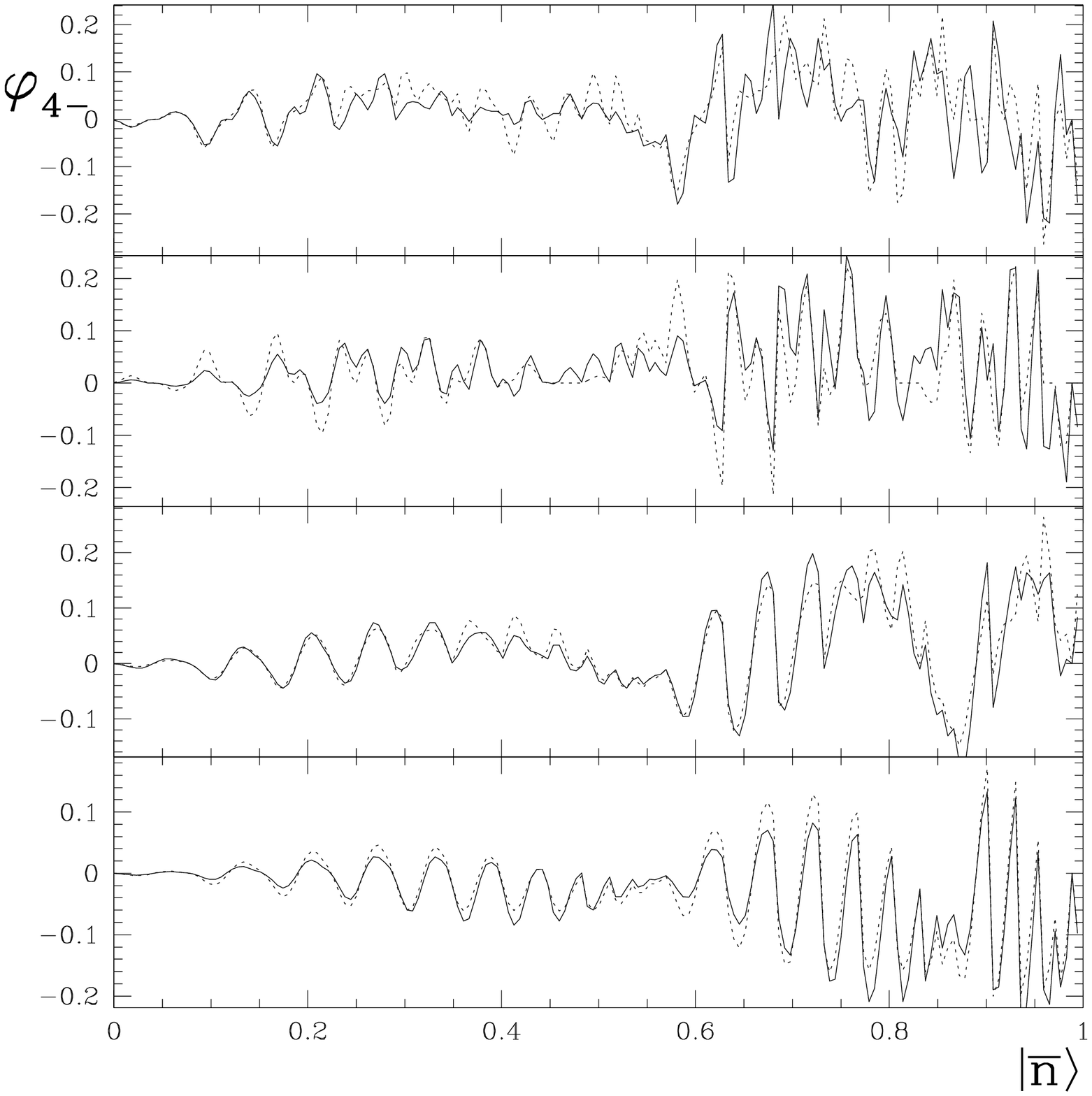,width=7.8cm}
}
\caption{Lowest eigenfunctions of the
  massive theory in the four-parton sectors.
Solid lines represent numerical results as emulated in DLCQ
by letting $\mu=4$, dashed lines the algebraic  
eigenfunctions.
Note that for $\mu=1$, the theory is supersymmetric. 
Left (a): ${\cal T}$ even eigenfunctions. Right (b): ${\cal T}$ odd
eigenfunctions.
\label{Fig4massivePartons}}
\end{figure}
%
%%%%%%%%%%%%%%%%%%%%%%%%%%%%%%%%%%%%%

What do the exact wavefunctions tell us? First off, the massless
and massive sectors show significant differences. The massive sector
is more straightforward, perhaps due to the more stringent constraint of 
wavefunctions vanishing on the $x_i=0$ hyperplanes. Even the lowest
states exhibit degenerate masses in the massive, but not the
massless theory. Also, the bound-state masses are much lower in the massless
sector. This is more surprising than it sounds, since we are omitting the mass
term in the asymptotic limit! The difference in bound-state mass is thus
generated {\em by symmetry alone}. The lowest four states are in the massive
theory ($\mu\neq 0$)
\bea
|1\rangle_{+-+}^{\mu\neq0} = |4,-2,0\rangle_{12}, \quad
&&
|1\rangle_{-+-}^{\mu\neq0} = |4,0,2\rangle_{12}, \quad \nonumber
\\
|2\rangle_{+-+}^{\mu\neq0} = |6,-2,0\rangle_{16}, \quad
&&
|2\rangle_{-+-}^{\mu\neq0} = |4,-2,0\rangle_{12}, \quad \nonumber
\\
|3\rangle_{+-+}^{\mu\neq0} = \frac{1}{\sqrt{2}}
\Big(|6,10,10\rangle_{20}+|8,10,6\rangle_{20}\Big), \quad
&&
|3\rangle_{-+-}^{\mu\neq0} = |6,4,6\rangle_{16}, \quad \nonumber
\\
|4\rangle_{+-+}^{\mu\neq0} = |8,10,10\rangle_{20}, \quad
&&
|4\rangle_{-+-}^{\mu\neq0} = |6,-2,0\rangle_{16}. \quad 
\eea
In the massless theory they look like
\bea\label{Massless4states}
|1\rangle_{+--}^{\mu=0} = |1,2,3\rangle_{6}, \quad
&&
|1\rangle_{-++}^{\mu=0} = |1,0,1\rangle_{4}, \nonumber
\\
|2\rangle_{+--}^{\mu=0} = |3,-2,-1\rangle_{10}, \quad
&&
|2\rangle_{-++}^{\mu=0} = |1,-2,-1\rangle_{6}, \nonumber
\\
|3\rangle_{+--}^{\mu=0} = |3,-2,-3\rangle_{12}, \quad
&&
|3\rangle_{-++}^{\mu=0} = |3,0,1\rangle_{8}, \nonumber
\\
|4\rangle_{+--}^{\mu=0} = |5,-2,-1\rangle_{14}, \quad
&&
|4\rangle_{-++}^{\mu=0} = |3,-2,-1\rangle_{10},
\eea
where the indices represent
the $TIS$ quantum numbers on the right-hand states,
and the mass (squared in units $g^2N\pi$) on the left-hand states. 
Note that the excitation numbers $n_i$ are not unique. For instance, at
$r=4$ each state has up to $2r!=48$ different tuples. One way of classifying
the state is to pick one statelet's numbers to represent the entire state,
e.g. by choosing the lowest positive number for $n_i$ followed by the smallest
absolutes $|n_i|$ as in Eq.(\ref{Massless4states}).

%%%%%%%%%%%%%%%%%%%%%%%%%%%%%%%%%% TABLE of States

\begin{table}
  \begin{footnotesize}
    \centerline{
\begin{tabular}{|c|lc|l|l|}\hline
$r$& T I S & Sector & Excitation numbers of lowest states & Masses ($g^2N\pi$)\\\hline
2 & $-+$ & $|o\rangle^0_+$ & $(1),(3),(5),(7)$ & $1,3,5,7$\\
  & $--$ & $|e\rangle^{\mu}_+$ & $(2),(4),(6),(8)$ & $2,4,6,8$\\
\hline
3 & $++$  & $|ee\rangle^{0}_-$ & $(0,0),(2,2),(4,2),(4,0),(6,2)$ & $0,4,8,8,12$\\
  & $--$  & $|ee\rangle^{0}_+$ & $(4,2),(6,2),(8,2),(8,4)$ & $8,12,16,16$\\
  & $+-$  & $|ee\rangle^{\mu}_-$ & $(2,0),(4,0),(6,2),(6,0)$ & $4,8,12,12$\\
  & $-+$  & $|ee\rangle^{\mu}_+$ & $(6,2),(8,2),(10,4),(10,2)$ & $12,16,20,20$\\
\hline
4 & $+--$  & $|oeo\rangle^{0}_+$ & $(1,2,3),(3,-2,-1),(3,-2,-3),(5,-2,-1)$ & $6,10,12,14$\\
  & $-++$  & $|oeo\rangle^{0}_-$ & $(1,0,1),(1,-2,-1), (3,0,1),(2,-2,-1)$ & $4,6,8,10$\\
  & $+-+$  & $|eee\rangle^{\mu}_+$ & $(4,-2,0),(6,-2,0),(6,10,10), (8,10,6), (8,10,10)$ & $12,16,20,20,20$\\
  & $-+-$  & $|eee\rangle^{\mu}_-$ & $(4,0,2),(4,-2,0), (6,4,6), (6,-2,0)$ & $12,12,16,16$\\
\hline
5 & $+++$  & $|eeee\rangle^{0}_+$ & $(0,0,0,0),(2,2,2,2),(2,4,4,4),(4,4,4,2),(4,4,4,4)$ & $0,4,8,8,8$\\
  & $---$  & $|eeee\rangle^{0}_-$ & $(4,4,4,2),(4,6,6,6),(4,6,4,2),(6,6,4,2),(4,8,8,8)$ & $8,12,12,12,16$\\
  & $++-$  & $|eeee\rangle^{\mu}_+$ & $(4,6,6,4),(6,8,8,6),(6,10,10,6),(8,10,10,6)$ & $12,16,20,20$\\
  & $--+$  & $|eeee\rangle^{\mu}_-$ & $(8,10,10,6),(8,12,10,6)$,(8,14,12,8) & $20,24,28$\\
\hline
6 & $-++$  & $|oeoeo\rangle^{0}_+$ & $(1,2,3,2,1),(1,2,3,4,3),(5,4,3,2,1),(3,4,5,4,3)$ & $6,8,10,10$\\
  & $+--$  & $|oeoeo\rangle^{0}_-$ & $(\ldots)$ & $$\\
  & $++-$  & $|eeeee\rangle^{\mu}_-$ & $$ & $$\\
  & $--+$  & $|eeeee\rangle^{\mu}_+$ & $$ & $$\\
\hline
\end{tabular}}
\end{footnotesize}
\caption{Characteristics of the lowest states in the first few parton sectors
  of the asymptotic theory including their quantum numbers $TIS$.
  %For $r=2,3$ we inferred the $S$ quantum numbers from operator redundancies.
  The sectors are labeled with a subscript
  indicating behavior under ${\cal T}$ ($_{\pm}$) and superscripts
  signifying massless ($^0$) and massive fermions ($^{\mu}$). For an explanation
  of the discrepancy between the $T$ quantum number and the ${\cal T}$ sector
  see Appx.~\ref{AppxSectorSigns}.
  \label{StateTable}}
\end{table}

%%%%%%%%%%%%%%%%%%%%%%%%%%%%%%%%%%%%%%%%%%%%%%%%%%%%%

\subsection{Some Group Theory}
\label{SecGroupTheory}

%Although we have the pragmatic goal of constructing a orthonormal
%harmonic basis of QCD$_{2A}$, we need to muster some group theory
%to see clearer why the solutions (\ref{TheSolution}) are so complicated.

%From a group theory perspective,
The solutions (\ref{TheSolution})
are right cosets of the subgroup ${\cal B}$ of
transformations associated with the full domain of integration
(the ``bulk'').
Namely ${\cal B}$ is the direct product of inversions ${\cal I}$,
reorientations ${\cal T}$
and cyclic permutations ${\cal C}$
\beq\label{Bsubgroup}
  {\cal B} = \left\{
  {1}, {\cal C}, {\cal C}^2, \ldots {\cal C}^{r-1},
  {\cal T}, {\cal TC},\ldots {\cal TC}^{r-1}, {\cal I}, {\cal IC}
  \ldots, {\cal ITC}^{r-1}
  \right\}.
  \eeq
We can construct a partition of the full group ${\cal G}$ of
symmetry transformations of QCD$_{2A}$ by acting on all elements
of ${\cal B}$
with the lower-dimensional inversions ${\cal S}$ concerning
symmetrization on the hyperplanes forming the boundary of the integration
domain
collected in the set\footnote{We refer to it as the exhaustive set ${\cal E}$,
  because its elements are exhausting the ``symmetrization space''.
  It is not a group, because it is not closed under composition of its members
  which may produce elements of ${\cal B}$.}
${\cal E}$ alluded to in Sec.~\ref{SecExhaustive}
\[
%  {\cal E} = \left\{
%  {1}, {\cal S}_1, {\cal S}_2, \ldots {\cal S}_{\bar{r}/2},
%  {\cal S}_1{\cal S}_2,{\cal S}_2{\cal S}_3,
%  \ldots {\cal S}_{\bar{r}/2-1}{\cal S}_{\bar{r}/2}
%    \right\}.
  {\cal E} = \left\{
  {\cal S}_1, {\cal S}_2, \ldots {\cal S}_{1/2(r-1)!-1}
    \right\}.
    \]
    In general, we find $N(r)=\frac{1}{2}(r-1)!-1$ independent
    lower-dimensional inversions.
    We work with the right cosets here, because we want to
    make explicit the symmetrization of the state under the cyclic
    subgroup $\langle {\cal C}\rangle$, as required by the structure
    of the Hamiltonian, as implicitly defined in Eq.~(\ref{TheEquation}).
    Since the order of ${\cal B}$ is
    $|{\cal B}|=2_{\cal T}\times 2_{\cal I}\times r_{\cal C}$ and its
    operators act on all elements of ${\cal E}$ plus the identity, a
    fully symmetrized state contains $2 r!$ ``statelets'' owing to the
    $2r!$ independent automorphisms that can be formed on a set of $r$
    objects (here: momentum fractions).

The general theme is hard to prove, so we simply
checked with a computer algorithm
that the order of ${\cal G}$ is indeed $|{\cal G}|=2r!$ ---
in particular, that it is {\em finite}.
%Of course, that is expected from
%the fact that in essence we are permuting $r$ ``things'', and from a group
%theory perspective it is unimportant
%that the ``things'' are parton momentum fractions. Indeed, from a mathematical
%point of view,
This is expected because permutations form
subgroups of the symmetric group. From a physical point
of view, permutations are re-orderings which leave the bound-state
masses invariant, and
thus symmetries of the Hamiltonian.

To check the order of ${\cal G}$ we proceeded as follows. By acting on all
$4r$ statelets ${\cal B}_i|r\rangle$ with the fundamental operator
${\cal S}={\cal S}_1$ we
produced the left coset ${\cal S}_1{\cal B}$. To keep the 
symmetry of the Hamiltonian manifest, we need however the {\rm right coset}
${\cal B}{\cal S}_1$. Recall that terms of odd $\cal C$-parity must be negative 
in the even $r$ sectors to ensure the alternating signs of the
permutations in the Hamiltonian
Eq.~(\ref{TheEquation}). Since in general the cosets are not {\em normal},
${\cal S}_1{\cal B}\neq {\cal B}{\cal S}_1$, we do not off-hand know which
${\cal C}$-parity the members of a left coset of ${\cal B}$ might have.
In general, both the left and the right coset will therefore contain
new operators, which we'll  have to include in our growing set of operators
to move towards
{\em group closure}. We must also act with these new operators on ${\cal B}$
as well as on the existing right cosets, which in turn will yield new operators,
right and left cosets. This process continues until no new operators arise
--- if the group is finite.
It is thus a non-trivial test of our hypothesis.
We find 
that the algorithm
always converges on the expected number of operators or statelets,
namely $2r!=240, 1440, 10080, 80640$ for $r=5,6,7,8$ and beyond.

To construct an actual eigenfunction, we need to solve one more problem.
It is the assignment of $S$ parity to the individual statelets, in the
sense that some of them will carry an odd power of $S$ quantum number, and
some an even power. Recall that $S^2=1$ since ${\cal S}$ is a $Z_2$ operation.
To do so it is convenient to introduce the notion of a
{\em primary operator}.
%, which we develop in the next section. 
%\subsection{Primary Operators and Conjugacy Classes}
Of the $\frac{1}{2}(r-1)!$ ${\cal S}$ operators in the $r$ parton sector,
initially only $r-1$ (associated with the $r-1$ excitation numbers) seem
necessary to ensure that the wavefunction is vanishing (maximal) on the
hyperplanes $x_i=0$ in the
massive (massless) theory, respectively. We call those $r-1$ operators
together with an additional operator\footnote{While not intuitive,
  the last operator is necessary to complete the map ${\cal S}_i\times
  \langle{\cal C}\rangle
  \rightarrow  \langle{\cal C}\rangle{\cal S}_i$, since the cyclic subgroup
  $\langle{\cal C}\rangle$ is of order $r$, not $r-1$.}
${\cal S}_r$ the {\em primary operators}.
They form a conjugacy class under $\langle {\cal C}\rangle$
and have similar properties, since\footnote{This relation
  is not unique, for instance
  ${\cal S}_{r-1} = ({\cal TC}){\cal S}({\cal TC})$ also.
  %Check if in general ${\cal S}_{r-i} = ({\cal TC}^i){\cal S}({\cal TC}^i)$
  }
\beq\label{GroupRelations1}
{\cal S}_i = {\cal C}^{i-1}{\cal S}{\cal C}^{r-i+1}.
\eeq
It is not hard to show that the primary operator fulfill the
following pseudo-commutation relations
\beq\label{pseudoCR}
{\cal C}^j{\cal S}_i={\cal S}_{{\rm mod}'_r (i+j)}{\cal C}^j\quad \mbox{and}\quad
({\cal TC}^j){\cal S}_i={\cal S}_{{\rm mod}'_r (r-i+j+1)}({\cal TC}^j),
\eeq
where ${\rm mod}'_r$ is congruence modulo $r$ shifted by one, so that
${\cal S}_0={\cal S}_r$ and ${\cal S}_{r+1}={\cal S}_1$.
In other words, the ${\cal S}$ operators commute with the
members of the normal subgroup $\cal B$
under loss of their identity: they transmute into a different primary
operator. But then we are done! These relations allow us to compute
all other operators in the symmetry group ${\cal G}$, which is in essence the
remainder of the subset $\cal E$ of lower-dimensional inversions. In fact,
we must keep adding operators until the $2r!$ slots in the full symmetry
group are {\em exhausted}. For instance, the product of two primary
operators can be evaluated as follows due to the associativity group
axiom\footnote{The $\cal T$ version of this relation is $
  {\cal S}_k({\cal TC}^j{\cal S}_i)=
  %({\cal S}_k{\cal TC}^j){\cal S}_i=
  {\cal TC}^j({\cal S}_{{\rm mod}'_r (r-k+j+1)}{\cal S}_i)= ({\cal S}_k{\cal S}_{{\rm mod}'_r (r-i+j+1)}){\cal TC}^j$.}
\beq\label{GroupRelations2}
  {\cal S}_k({\cal C}^j{\cal S}_i)=({\cal S}_k{\cal C}^j){\cal S}_i=
  {\cal C}^j({\cal S}_{{\rm mod}'_r k-j}{\cal S}_i)=
  ({\cal S}_k{\cal S}_{{\rm mod}'_ri+j}){\cal C}^j\quad
  \quad\mbox{for}\quad 1 \le i,j,k \le r.
\eeq
The second half is the commutation relation for the {\em secondary}
operator ${\cal S}_{{\rm mod}'_r(k-j)}{\cal S}_i$. 
A complete multiplication table of the group can
be iteratively constructed.
Finding independent operators essentially
reduces to the {\em word problem} of abstract algebra.
We display part of such a table for $r=5$ in
Table \ref{GroupMultTable_r5}.
%It is easy to show that in general\footnote{That is, for $r>4$.}
%the highest power operator is a product of $r-1$ primary operators. 
As a corollary we note that these identities show that the Hamiltonian
symmetrization constraint (fixed signs under $\langle {\cal C}\rangle$) is
intact.
%--- a necessary condition for our method to work!
Namely, the ${\cal T}^i{\cal C}^j$ component of a right coset of ${\cal B}$
of any ${\cal S}_k$ (whether primary or not) is mapped onto the
${\cal T}^i{\cal C}^j$ component of a right coset of some other operator
${\cal S}_{k'}$.

%Still, we are not quite ready to construct generic states.
So what are the relative signs of the individual terms? The signs
{\em within}
the $\cal B$-blocks ($4r$ statelets of the right cosets ${\cal BS}_i$)
are fixed by the Hamiltonian and the
$T$ and $I$ quantum numbers.
%What is the additional multiplicative $S$ quantum number of the blocks? 
At $r=4$ the assignment of the $S$ quantum number is simple
if counter-intuitive: both the ${\cal S}_1$ and the
${\cal S}_2$ block 
carry an additional $S$ sign (32 statelets); only the identity block's
eight statelets do not acquire a sign. This seems unbalanced, but works
due to sign-cancellations unique to the four-parton sector.
At $r=5$ the orders of group and
subgroups are such that we get an even split into two ``half-groups''. Namely,
the five secondary operators plus the identity are $S$ even
and the primary operators plus a peculiar secondary operator\footnote{It commutes with the $\cal B$ subgroup,
i.e. ${\cal B}S_9=S_9{\cal B}$, which does not imply commutation of
  individual subgroup elements ${\cal B}_i$ with ${\cal S}_9$: 
  ${\cal B}_iS_9\neq S_9{\cal B}_i$ in general.} ${\cal S}_9$ are odd and
carry an $S$ sign.  This makes $r!=120$ statelets with an
$S$ sign, and $120$ without.
%This cannot work at $r=6$, since $2r!=1440$ is not
%divisible by $4r=24$.
For an algorithm that is more universally applicable, see Appx.~\ref{DerivationOfGMT}.

%It is clear that the $r$ primary right cosets carry an $S$ sign. Logic tells
%us then that all odd operators (primary, tertiary, quinary, etc.) should
%carry a $S$ sign, and the even operators should not. We might call this
%the $S$-{\em parity} of the operator. 
%This hypothesis
%works out at $r=4$, and at $r=5$ the peculiar operator ${\cal S}_9$ is indeed
%tertiary. But are all others even? Let's check: ${\cal S}_{11}$ is quarternary
%(highest power!), ${\cal S}_{6}={\cal S}_{1}{\cal S}_{2}$,
%${\cal S}_{7}={\cal S}_{1}{\cal S}_{3}$,
%${\cal S}_{8}={\cal S}_{2}{\cal S}_{3}$,
%${\cal S}_{10}={\cal S}_{3}{\cal S}_{2}$ are all secondary as required,
%and ${\cal S}_{11}={\cal S}_{5}{\cal S}_{2}{\cal S}_{1}{\cal S}_{3}$ is
%quarternary, i.e.~even.
%Hence, the algorithm is nothing but a classification of emerging operators
%as odd or even power of primary ${\cal S}$ operators. 

%What remains to be shown is that for the non-primary operators all
%members of a right coset have the same $S$-parity, and that acting with
%a primary operator on a right coset changes the coset's $S$-parity.
%We need to check this, since at $r=6$ a peculiar operator,
%(${\cal S}_{48}={\cal S}_{5}{\cal S}_{3}{\cal S}_{2}{\cal S}_{3}$),
%is nominally quarternary, but appears in both even and odd
%right cosets. Is this behavior due to the fact that two of its components
%are identical? But if they cancel, it is still an even operator. In
%any case, it is a counter-example so far. (SHOW THAT!)

\begin{figure}
\centerline{
\psfig{file=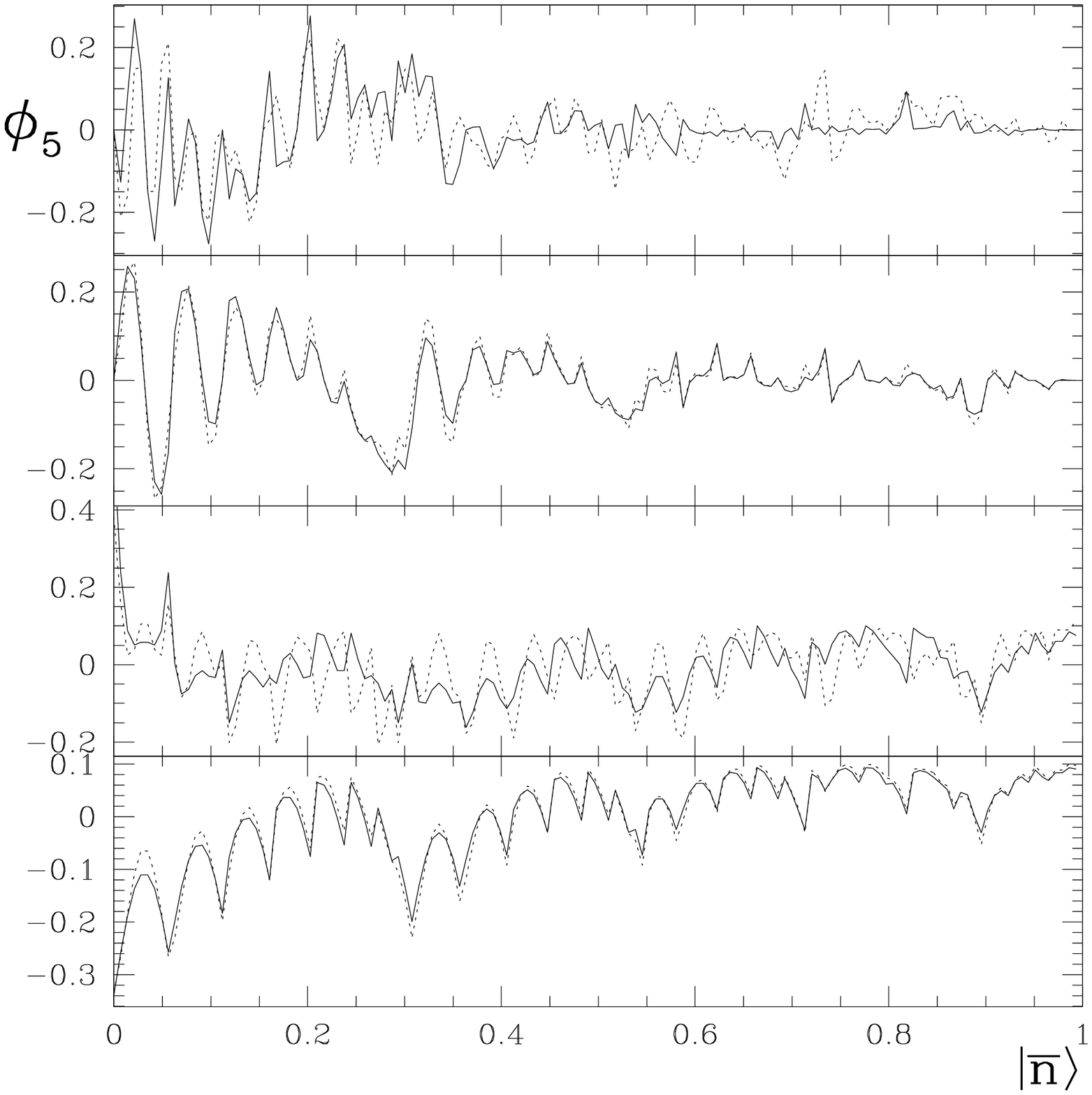,width=7.8cm}
\psfig{file=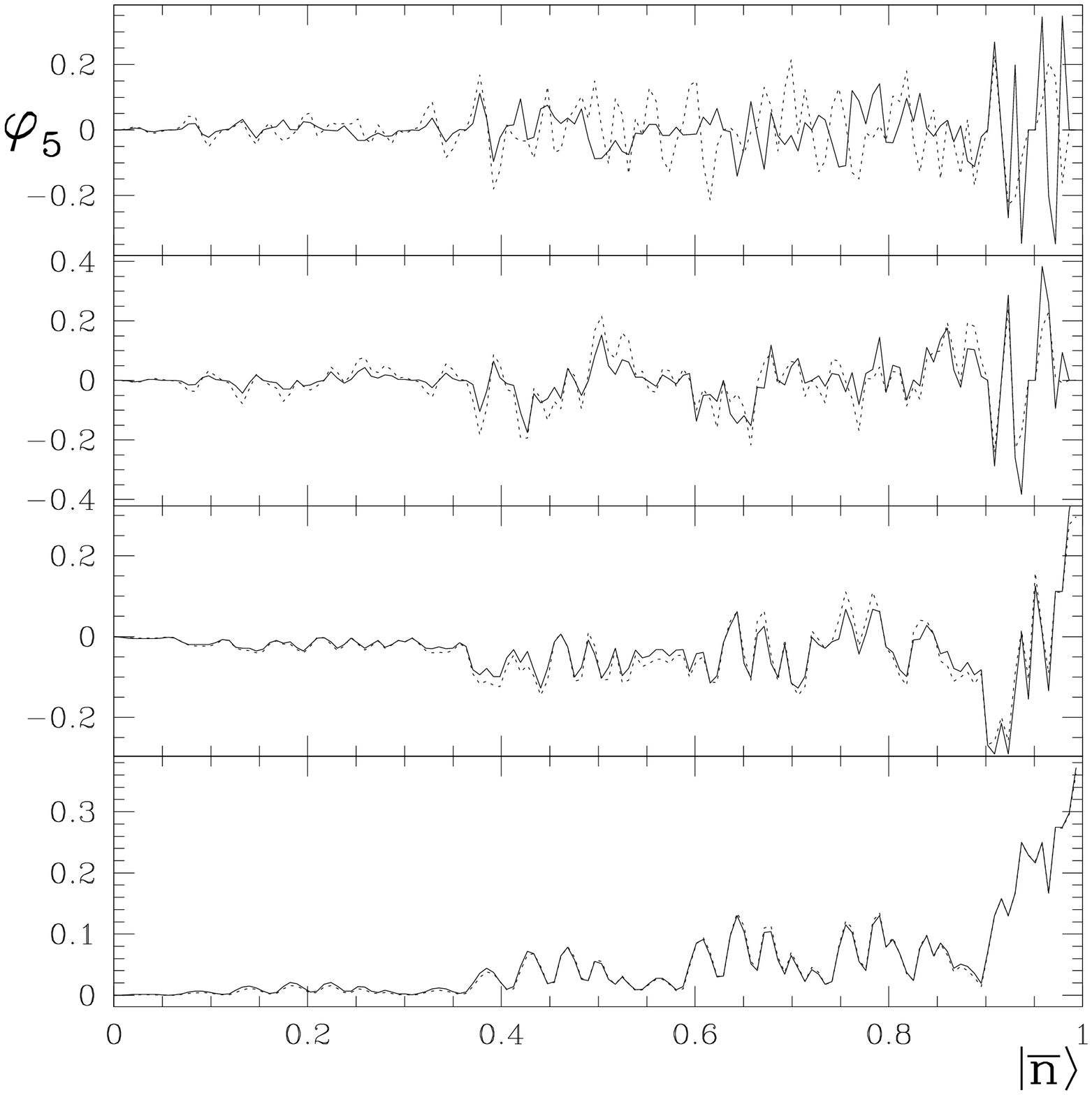,width=7.8cm}
}\caption{Lowest eigenfunctions in all five-parton sectors.
  Solid lines represent numerical (DLCQ) results, dashed lines the algebraic  
eigenfunctions.
Left (a): Massless theory ($\mu=0$). Right (b): Massive theory ($\mu=4$).
The lowest two eigenfunctions are ${\cal T}$ even, the upper two ${\cal T}$
odd. 
\label{Fig5partons}}
\end{figure}

This completes our construction of the most general solution in all
parton sectors. As a cross-check we construct the eigenfunctions in
the five-parton sector and compare them to numerical
solutions in Fig.~\ref{Fig5partons}. 
Fig.~\ref{KM_WFs_RealAndModel}(a)
shows the comparison for some six-parton eigenfunctions.
We used the rather low harmonic resolutions $K=23$ and $K=20$,
because otherwise the plots get too ''crowded''; agreement is just as
good at higher $K$ of course.  As we glean from Fig.~\ref{Fig5partons}, the
agreement is near-perfect for the lowest ${\cal T}$ even state in the massive
theory, good in the massive theory in general, and fair in the
massless theory. 
This might be due to the fact that the density of
states is much higher in the massless theory. Apparently, the
condition that the wavefunction vanish for zero parton momenta is
quite restrictive. Given that the number of statelets rises to 1440 at $r=6$
the agreement between
algebraic eigenfunctions and numerical solution seems surprisingly
good in Fig.~\ref{KM_WFs_RealAndModel}(a).

It is indeed quite remarkable how perfectly the features of the theory are
represented by the properties of the harmonic basis. It is tempting to
speculate
%that the harmonic functions constitute a faithful
%representation of the theory in function space and
that this approach
--- which one might call ``{\em exhaustively-symmetrized Light-Cone Quantization}
(eLCQ)'' 
%since it was borne out of the spirit of DLCQ
--- applies to other
theories, at least in two dimensions where light-cone coordinates
are the natural language. Hopefully, the treatment of the present toy
model is a harbinger of wider applicability.
%With this in
%mind, we will try to explain as many of the salient features of the
%full version of QCD$_{2A}$ features of the basis functions in the
%next sections and shall suggest other viable theories in the
%discussion.

%%%%%%%%%%%%%%%%%%%% Massless & Massive Five %%%%%%%%%%%%%%%%%
%
%
%%%%%%%%%%%%%%%%%%%%%%%%%%%%%%%%%%%%%

%---------------------------------------------
\section{Applications}
\label{SecApplications}
%---------------------------------------------

\subsection{Using ``eLCQ'': Some Suggestions}
%\label{SecAccidental}
\label{SecUsing_eLCQ}

%------------------------------------------------------------
%\subsection{Approximating the Full Theory: A Basis-Function Solution of QCD$_{2A}$}
%\label{FullTheory}
%------------------------------------------------------------

Since we have found a complete set of basis states for the asymptotic
theory in Sec.~\ref{SecBasis}, we can systematically approximate
the full theory in a basis-function approach.
This should be pretty straightforward, but implementing
the algorithm in the higher parton-sectors is too tedious to be presented
here, so we only sketch the general idea.
%and leave the details for
%later \cite{UTFututeWork}.
Namely, we can use the set of coupled integral equations
derived in \cite{BDK} from the Hamiltonian Eq.~(\ref{P_BDK}) 
for the full wavefunctions $f_r$ in the
$r$-parton sectors. The full
equation couples sectors of different parton number;
$f_r$ and $f_{r\pm 2}$ appear in the equation.

We expand the full eigenfunctions $f_{r}(x_1,x_2,\ldots, x_r)$
into a complete set of asymptotic eigenfunctions $\phi_{r,\vec{n}}$
\[
f_{r}(x_1,x_2,\ldots, x_r)=\sum_{\vec{n}}c_{r,\vec{n}}\,
\phi_{r,\vec{n}}(x_1,x_2,\ldots, x_r),
\]
where
$\vec{n}$ represents a tuple of $r-1$ excitation numbers.
Formally, the integral equation looks like
\[
M^2f_r = M^2\sum_{\vec{n}}c_{r,\vec{n}}\phi_{r,\vec{n}}(x_1,x_2,\ldots, x_r) = 2P^+P^-
\sum_{\vec{n}}c_{r,\vec{n}}\phi_{r,\vec{n}}(x_1,x_2,\ldots, x_r).
\]
If we project onto the asymptotic eigenfunctions characterized by $(s,\vec{m})$,
we get an equation for the associated coefficient
\[
M^2\int d^rx f^*_{s,\vec{m}}(\vec{x}) f_r(\vec{x}) =  M^2\sum_{r,\vec{n}} \int d^rx c_{r,\vec{n}}
f^*_{s,\vec{m}} f_{r,\vec{n}} = M^2  \sum_{r,\vec{n}} c_{r,\vec{n}}\delta_{s,r}\delta_{\vec{n},\vec{m}} = M^2 c_{s,\vec{m}},
\]
where $\vec{x}$ represents the $r$ momentum fractions $x_i$, and
$\int d^rx$ the integration over the appropriate domain subject to
the constraint $\sum_i x_i=1$ {\em and} the removal of cyclic redundancies,
cf.~Eq.~(\ref{TheStates}). We assumed
the asymptotic eigenfunctions to be orthonormal and complete.
On the left hand side, we have to evaluate the Hamiltonian matrix elements
and multiply the matrix with the column vector of coefficients. In other
words, we have to solve an eigenvalue problem for the coefficient vectors.
Diagonalizing the full Hamiltonian clearly
will yield the coefficients to express
the full eigenfunctions as linear combinations of the asymptotic
eigenfunctions.

%Accidental Symmetries
%\label{SecAccidental}

Now, the number of statelets in a state is equal to the order
of the group $|{\cal G}|=2r!$ and grows exponentially.
This limits the practical value of the approach. However, 
it is in some sense the worst case scenario.
Often, accidental symmetries arise due to
special combinations of excitations numbers $n_i$. For instance, the
four-parton eigenfunctions, Eq.~(4.13) in Ref.~\cite{Kutasov94}, can be
generated by symmetrizing the statelet $|n_1,0,n_2\rangle$. Most of
its statelets will not exhibit a vanishing excitation number, yet the
state as a whole is more symmetric than the generic four-parton
eigenstate. To wit, it possesses an additional $Z_2$
symmetry\footnote{Its identity and ${\cal S}_1$ statelets fulfill
  ${\cal TZ}|r\rangle={\cal Z}|r\rangle$, whereas for the
  ${\cal S}_2$ statelets we have ${\cal C}^2|r\rangle={\cal I}|r\rangle$,
  so that half the statelets are redundant.}, so that the 24 independent
statelets can be cast into 12 sinusoidal functions, or into the 6 double-sines
of  Eq.~(4.13) in Ref.~\cite{Kutasov94}. Analogous
symmetries seem to exist in all higher parton-sectors, e.g.~six-parton
states $|n_1,0,n_2,0,n_3\rangle$
generating states\footnote{Note that this works in the
  massless $|oeoeo\rangle$ sector, too.}
of the form Eq.~(4.15) in Ref.~\cite{Kutasov94}.
All symmetries will reduce the eigenvalue problem further by
block-diagonalizing the Hamiltonian and may
give an intuitive understanding of the bosonization process, by which
trivial multi-particle states are projected out \cite{KutasovSchwimmer}.

%--------- S operations in momentum space

So far the emphasis has been on symmetries of the set of excitation numbers.
This is natural, since the ${\cal S}$ symmetries were introduced exactly
for this purpose in Eq.~(\ref{LDI}). A wavefunction symmetrized with
respect to its excitation numbers is clearly also symmetrized with respect to
its arguments, i.e.~momentum fractions. What kind of momentum space symmetries
do the $N(r)$ ${\cal S}$ operators represent? It should be enough to explicitly
look at only two, since all others can be derived from them. Namely,
${\cal S}_1$ and ${\cal S}_2$ can be considered
stereotypical single- and double-neighbor permutations, respectively,
since in the latter case $n_2\rightarrow n_2-n_1-n_3$ is affected by both of the
neighboring excitation numbers.
It is clear that the shift in excitation numbers comes from the following
shift in momentum fractions 
\bea
    {\cal S}_1: && x_1\rightarrow -x_1-x_2, x_2\rightarrow x_2, x_3=-x_3-x_1, \ldots ,
    x_{r-1}\rightarrow -x_{r-1}, x_r\rightarrow -x_r-x_1,\nonumber \\
    {\cal S}_2: && x_1\rightarrow x_1, x_2\rightarrow -x_2-x_1, x_3=-x_3, \ldots
    x_r\rightarrow -x_r\nonumber. 
\eea
Apparently there is only one operation in momentum space:
all (except one) momentum fractions are inverted, and the fractions
next to the invariant one are shifted by the invariant fraction. 
This shows that the condition that the Hamiltonian be hermitian, and hence
that its eigenfunctions are invariant under ${\cal S}$, generates a set
of momentum space symmetries that grows exponentially with the parton number.
Since these generated symmetries are not obvious, the general method may be useful to determine a full set of symmetries of a given Hamiltonian in
other theories. 

%------------------------------------------------------------
\subsection{The Role of Pair Production}
\label{SecPairProduction}
%------------------------------------------------------------

With the asymptotic solution Eq.~(\ref{TheSolution}) at hand, what
can we say about the effect of pair production, i.e.~parton number violation?
The first observation is that the subsequent parton sectors with
the same ${\cal T}$ parity have opposite ${\cal I}$ parity. In other
words, the Hamiltonian is sandwiched between a cosine and a sine
wavefunction.
The simplest case is the matrix element between the two and four
massless parton ${\cal T}$ even sectors\footnote{The subscripts here are the $TI(S)$ quantum numbers, consistent with Table \ref{StateTable}.}
\[
_{-+}\langle r=2; 1 |P^-_{PV}|r=4; 1\rangle_{+--},
\]
where
\bea
|r=2; 1\rangle_{-+}&=& \cos \pi x ,\nonumber\\
|r=4; 1\rangle_{+--}&=&
\sin\pi ( x_1+2 x_2+3 x_3)-\sin\pi (3 x_1+2 x_2+ x_3)\nonumber\\
&&+\sin\pi ( x_1-2 x_2- x_3)+\sin\pi ( x_1+2 x_2- x_3)\nonumber,
\eea
and the latter wavefunction has an interesting ``disappearance'' symmetry
under $x_2\leftrightarrow x_4$.
In general, we will have to evaluate a matrix element of the
form
\[
_{-+}\langle \bar{n} |P^-_{PV}| n,m,l\rangle_{+--},
\]
where $\bar{n}$ and $n,m,l$ represent the excitation numbers
of the two- and four-parton states, respectively, and $P^-_{PV}$ is defined in
Eqs.~(\ref{StructureOfHamiltonian}) and (\ref{P_BDK}). Owing to the definition
of Hamiltonian and states, Eqs.~(\ref{P_BDK}) and (\ref{TheStates}), we have to
do eight integrations over momenta (one plus three for the states, and four for
the Hamiltonian), and have one momentum conserving delta-function, as well
as five more appearing when we commute through annihilation operators. So we are left with two integrations over a trigonometric function divided by a quadratic
function of the momenta. The structure of the final expression is therefore
\[
_{-+}\langle \bar{n} |P^-_{PV}| n,m,l\rangle_{+--}
\sim \int\int \frac{\sin\pi(n'x+m'y)}{(x+y)^2}dxdy,
\]
where the integers $n'$ and $m'$ will depend on all excitation numbers
$\bar{n},n,m,l$ of the two and four-parton states. This integral can
be expressed in terms of cosine and sine integrals ($Ci(x), Si(x)$), and
is not divergent. There will be several terms, and cancellations are possible.
Thus
for specific states, or in certain sectors\footnote{Of course,
  the ${\cal T}$ sector is fixed, but the result can be different
  in the massive and
  massless sectors of the theory.} the matrix element may vanish.

This in turn raises the question whether it is systematically
possible to find linear
combinations of the asymptotic eigenstates for which the annihilation matrix
elements are zero. These would then be states with a definite parton number.
In other words, we would have succeeded in summing all pair creation effects,
effectively ``renormalizing'' the theory, i.e. formulating it in terms of
effective degrees of freedom. This program is beyond the scope of the present
note, but it seems to be viable --- judging from previous work.
Namely, in \cite{DalleyKlebanov,BDK} it was shown that
many of the bound states are very pure in parton number.
An open question is whether this behavior
is generic or due to the finite discretization used in
\cite{DalleyKlebanov,BDK}.

%------------------------------------------------------------
\subsection{Implications for the Bosonized Theory}
\label{SecBosonized}
%------------------------------------------------------------

QCD$_{2A}$ can be bosonized by rewriting the Hamiltonian in terms of
current operators $J(-p)\sim\int dq\, b(q)b(p-q)$ subject to a Kac-Moody
algebra, see e.g.~Refs.~\cite{Adi95, UT}. Bosonization is
in essence a basis transformation, so the eigenfunctions will change
while the eigenvalues, i.e. the bound state masses, remain invariant.
Straightforward bosonization generates a non-ortho\-normal basis, with
states consisting of color traces of adjoint {\em current operators}
acting on a vacuum state\footnote{Either the traditional vacuum state
  or an ``adjoint vacuum'', represented by a fermionic operator of
  zero momentum acting on the vacuum.}.
Numerical approaches \cite{UT} produce orthonormal solutions, i.e.~the
coordinate vectors associated with a chosen basis (whether
orthonormal or not) are mutually orthogonal and of unit length.
In this sense a basis of symmetrized harmonics as produced by
our ``eLCQ'' method will be orthonormal.

\begin{figure}
\centerline{
  \psfig{file=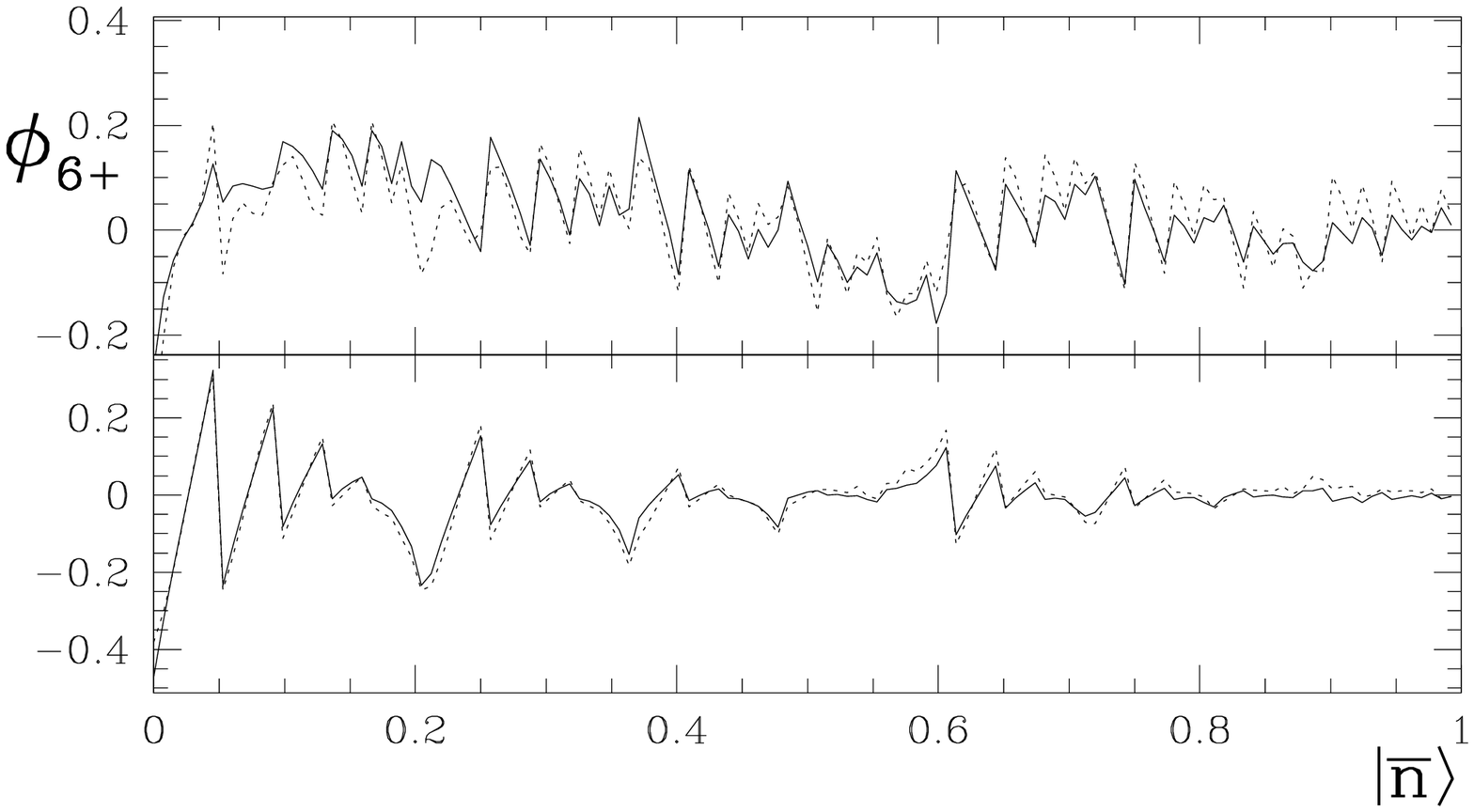,width=7.8cm}
  \psfig{file=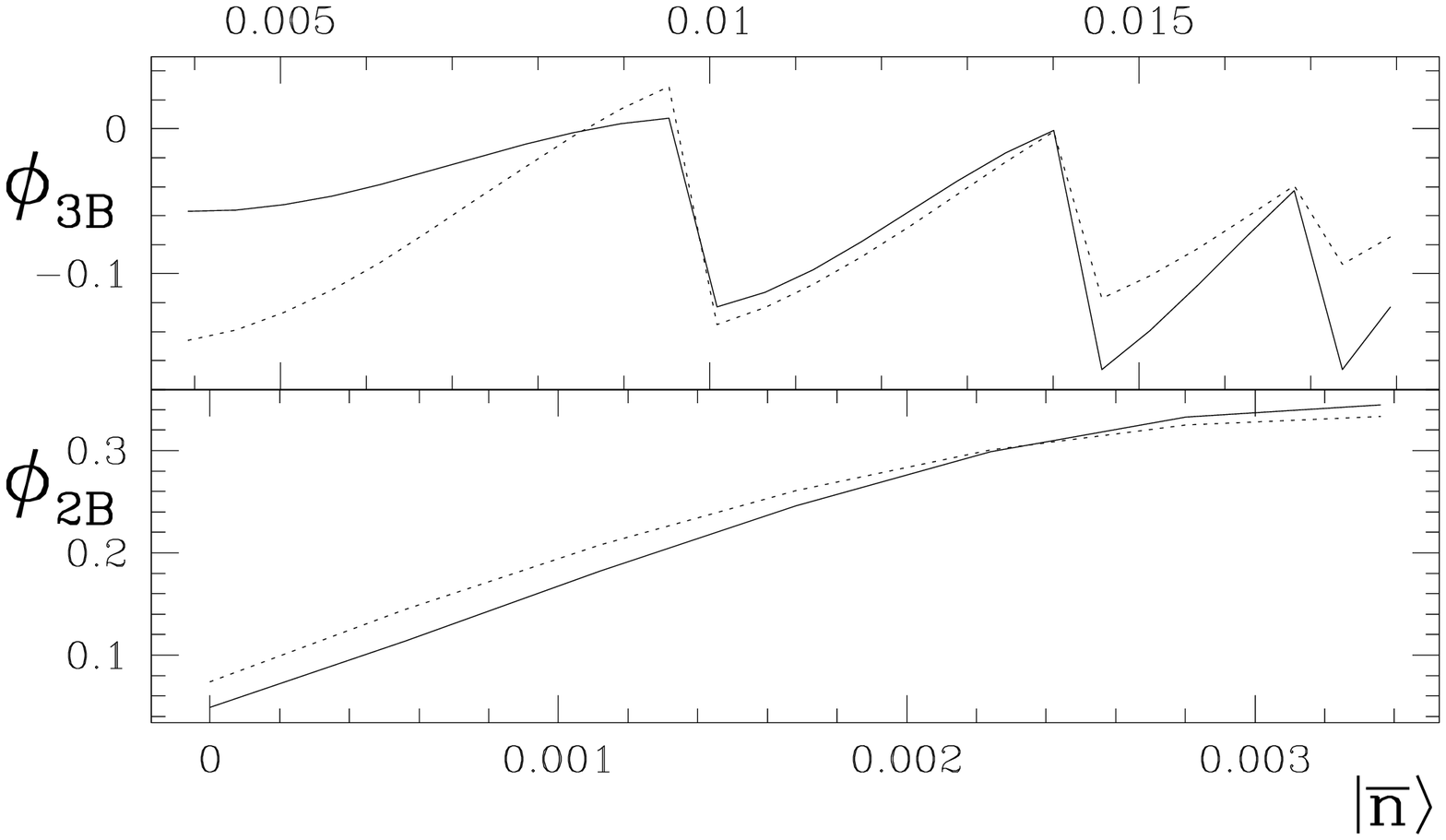,width=7.8cm}
}
\caption{(a) Left: Massless six-parton ${\cal T}$ even
  eigenfunctions of asymptotic $QCD_{2A}$ at $K=20$.
  (b) Right: The two- and three-parton parts of the
  the lowest ${\cal T}$ even eigenfunction of the {\em bosonized theory} at
  $K_B=14$ which is equivalent to $K=28$ in the fermionic theory. 
  In both graphs DLCQ eigenfunctions are plotted with solid lines and
  ``eLCQ'' or model wavefunctions with dashed lines. 
\label{KM_WFs_RealAndModel}}
\end{figure}

To get a handle on the bosonized theory, current-number changing operators can be omitted at first \cite{Adi95}, and
one arrives at an integral equation for the bosonized eigenfunctions $\phi_{rB}$
\beq
\frac{M^2}{g^2N}\phi_{rB}(x_1,\ldots,x_r)=-\sum_{i=1}^r
\int_{-\infty}^{\infty}\frac{\phi_{rB}(y,x_i+x_{i+1}-y,
x_{i+2},\ldots,x_{i+r-1})}{(x_i-y)^2}dy,
\label{TheBosonizedEquation}
\eeq
where we
have used the subscript $B$ to stress that the partons in the bosonized
theory are currents and not fermions. The equation is virtually 
identical with Eq.~(\ref{TheEquation}) save for the non-alternating signs. 
These signs stemming from anti-commuting fermionic operators are clearly  
absent in the bosonized theory. We should therefore be able
to describe the bosonized wavefunctions with our ``eLCQ'' ansatz,
Eq.~(\ref{TheSolution}). We cannot hope for a perfect match, because we
cannot decouple sectors of different current number. A consistent
asymptotic theory does not exist due to the nature of the 
Kac-Moody algebra
of the current operators. There is an associated problem. Namely,
the full theory contains uninteresting
non-trivial multi-particle states which interact
with the single-particle states of interest at any finite resolution
in discretized versions of the theory \cite{GHK,UT3}.
Indeed, only a few single-particle states have hitherto been identified as
such. Furthermore, bosonization only works for massless
fermions, so there is no massive sector. Hence, our comparison of
numerical and algebraic wavefunctions will be rather limited.
Of course, in the bosonized theory there is a
bosonic and also a fermionic sector. In the latter, the basis states are different owing to the elimination of cyclic symmetry due to the existence of a
unique fermionic operator (the ``adjoint vacuum'').
Here, we will focus on the bosonic sector
of the bosonized theory. This is a real test of our ansatz, since
the symmetrization clearly cannot be the same as in the theory with fermions.
The ``eLCQ'' ansatz (\ref{TheSolution}) for
the two-current wavefunction $\phi_{2B}$ is
\[
\phi_{2B}=e^{inx}+(-1)^n e^{-inx}
\]
(compare to Eq.~(\ref{TwoPartonState}) which has a minus sign between
the terms). Therefore we ``predict'' that the massless states with odd $n$
will be sines in the bosonized theory, not cosines as in the fermionic
theory. This is consistent with Ref.~\cite{Adi95}, and also
the numerical bosonized eigenfunctions in
Fig.~\ref{KM_WFs_RealAndModel}(b) are
consistent with our ansatz\footnote{We used
$\phi_{2B}=\sin(\pi x_1)$  and 
$\phi_{3B}=-\frac{1}{20}
  \left[\cos(\pi x_1)+\cos(\pi x_2)+\cos(\pi x_3)\right]$ to describe the
  generic features of the two wavefunctions. Note that the cosines in the
  higher parton sectors are consistent with our earlier finding that
  wavefunctions of same $T$ have opposite $I$ at subsequent $r$.}.
Note that numerical approaches such as DLCQ use different bases
for the fermionic and bosonized theories.
The fermionic theory is approximated using anti-periodic boundary conditions
(odd half-integer momentum fractions), whereas the bosonized theory
uses periodic boundary conditions (integer momentum fractions). Of course,
regardless of the basis used, they should be decent approximations to
the algebraic wavefunction.

%Unfortunately, it looks like this does not work out. The wavefunctions
%of the fermionic theory's three and four parton sectors look nothing like the
%bosonized wavefunctions. This is very strange, since to zeroth approximation
%they obey the same integral equation.

In sum, we find that the ``eLCQ'' ansatz is compatible with the known solutions
of the theory in a significantly different representation. This hints at a
wider applicability of our method.
What seems crucial is that the structure of the Hamiltonian be of the form 
\[
P^-\sim \int \frac{dp}{p^2}J_{ij}(-p)J_{ji}(p).
\]
In other words, the ``eLCQ''
ansatz is poised to solve the long-range Coulomb-type
part of a strongly interacting system. This is akin to the DLCQ ansatz which
was shown in Ref.~\cite{Pauli84} to decouple the center of mass motion of
a system from its more interesting physics. 

%%%%%%%%%%%%%%%%%%%% r=6 and KM_WFs_RealAndModel %%%%%%%%%%%%%%%%%
%
%
%%%%%%%%%%%%%%%%%%%%%%%%%%%%%%%%%%%%%

%------------------------------------------------------------
\section{Conclusion and Discussion}
\label{SecDiscussion}
%------------------------------------------------------------

The goal of this paper was to find a complete set of eigenfunctions of
QCD$_{2A}$. We succeeded in constructing a basis of the asymptotic
theory without pair-production, consisting of multi-dimensional
harmonic functions.  In order to completely and
exhaustively symmetrize the wavefunctions in the $r$ parton sector, a
{\em group} of operators of order 2r!  is necessary and
sufficient. This finite group of abstract symmetries 
is defined by the relations between the generators
Eqs.~(\ref{GroupRelations1})--(\ref{GroupRelations2}). Paradoxically,
this rather complicated arithmetic is based on
the simple observation that ``boundary conditions'' on the wavefunctions
have to be implemented as symmetries in a Hamiltonian approach (which
leads to an integral equation for the eigenfunctions).  Our finding
that the eigenfunctions can be constructed largely algebraically with
group theory arguments is corroborated by comparison with numerical
solutions of the theory.

%Since all finite groups have been classified, we have thus complete
%knowledge of this group.

 At first glance our method of exhaustively symmetrizing a system quantized
 on the light-cone (``eLCQ'') applies to a 
specific family of systems, namely theories with
adjoint degrees of freedom in one spatial dimension with a Coulomb-type
long-range interaction. It seems likely, though, that the method presented here
is more widely applicable due to its generality.
%Indeed, the approach is not as necessarily confined to two-dimensional
%theories as it may seem. The advantage of light-cone quantization
%in two dimensions is the compactness of the momentum domain.
%Here, however, it has presented us with difficulties, as we had to formulate
%our eigenfunction ansatz with $r-1$ quantum numbers, since the $r$ longitudinal
%momentum fractions $x_i$ add up to unity. But nothing hinges on this
%construction. We can symmetrize any set of quantum numbers. In a sense, that's
%what we have done by considering all parton sectors $r$. The other
%argument that exhaustive symmetrization might not work for transverse degrees
%of freedom is that the momentum integral boundaries are pushed to infinity,
%since transverse momentum is not cut off. Hence, the argument goes, we cannot
%imposes boundary conditions (other than that the wavefunction has to
%vanish). But we do not wish to! We reiterate that
%the ``eLCQ'' approach produces rather than specifies boundary conditions.
%By symmetrizing the wavefunction we achieve specific
%behavior at the boundaries as a by-product. Also
In particular, exhaustive
symmetrization deals with momentum as an abstract entity, i.e. it deals
with it regardless of its Lorentz structure. There might well be requirements
due to the Poincar\'e group, but this is not the point here. Rather, the
fact that we have $r$ momenta requires us to symmetrize on purely
abstract grounds, rather as a preconditioning of the
wavefunctions akin to the Slater determinant implementing the
anti-symmetrization requirement of the Pauli principle.
%In that respect, it might be interesting to rethink
%the Coleman-Mandula theorem as it applies to theories with
%exhaustive symmetry; the latter might be classifyable as an internal
%symmetry.(THIS MIGHT be NONSENSE, so check!) 

We presented evidence that ``eLCQ'' is
a useful method at least for theories exhibiting similar
integral equations as QCD$_{2A}$ like its bosonized
version as discussed in Sec.~\ref{SecBosonized}.
Other theories to which ``eLCQ'' can be straightforwardly
applied include two-dimensional Yang-Mills
theory coupled to adjoint scalars \cite{DalleyKlebanov} and a theory
with adjoint Dirac fermions tackled in \cite{GHKSS}. Also, many
of the supersymmetric models with adjoint particles (for instance
\cite{HillerPinskyTrittmann}) could be re-evaluated with the present method.

We did not have the space to fully exploit the fact that the
asymptotic spectrum of
QCD$_{2A}$ is now completely mapped out, so there are several opportunities
for future projects.  For instance,
the interaction of asymptotic states and the
formation of multi-particle states can be studied. In
Sec.~\ref{SecPairProduction} we laid down an initial plan how to
proceed. And while the exponential rise of terms in the eigenfunctions
presents some difficulty, one might be able to improve the precision of
numerical solutions enough by using ``eLCQ'' basis functions to
positively identify the single-particle content of the
theory. Accidental symmetries described in Sec.~\ref{SecUsing_eLCQ}
will help in this regard.  Also, the structure of the asymptotic
solutions may help to disentangle salient features of this and other
theories.  For instance, the ``topological sector'' of the theory,
introduced in \cite{GHK} to explain the appearance of fermion-fermion
multi-particle states in the {\em fermionic} sectors, might be an
artifact of finite group theory. As long as parton number is finite,
there is a clear separation of group properties according to their
(finite) order.  Some combination rules for multi-particle
states in terms of $T$ quantum numbers found in \cite{GHK} might just
be the result of such ``artificial'' group theoretical relations.

%------------------------------------------------------------
\section*{Acknowledgments}
%------------------------------------------------------------

I thank Otterbein University for providing a work environment which made
it possible to see this long-term project through to completion. 
The hospitality of the Ohio State University's Physics 
Department, where most of this work was completed, is gratefully acknowledged.

The method of exhaustive symmetrization was conceived
ten years after Hans-Christian Pauli's passing. I'd
like to dedicate the present paper, which I believe was done in the spirit
of his work starting with \cite{Pauli84}, to his memory.

\begin{appendix}

\section{Derivation of the Group Multiplication Table}
\label{DerivationOfGMT}

To fully understand why there is only one quantum number $S$ associated
with the large subgroup ${\cal E}$ of ${\cal S}$ operators, we work
out the details here.
We start with the simplest case $r=4$.
%--- which is unfortunately not generic.
We need to show that the fully symmetrized wavefunction ${\cal G}|r\rangle$
is an eigenstate to all three ${\cal S}$ operators associated with its
three excitations numbers $(n,m,l)$. Note that {\em de facto} there
are only {\em two} independent ${\cal S}$ operators at $r=4$, which we
also have to explain, along with the fact that two thirds --- not half ---
of a state's statelets carry an $S$ sign.

First we observe that the $4r$ members of the subgroup ${\cal B}$ naturally
split into four subsets, namely the cosets of the cyclic subgroup
$\langle{\cal C}\rangle$ under
${\cal T}$, ${\cal I}$, ${\cal T}$, and  ${\cal IT}$. 
But ${\cal I}$ is in the center of ${\cal G}$, and therefore can be largely
ignored. The cyclic structure of the Hamiltonian means that there
is a conserved quantity which we might call ${\cal TC}$-parity.
Odd and even powers of the cyclic group carry different signs in general.
Therefore an even(odd) power has to be mapped under any ${\cal S}$
operation onto an even(odd) power of of ${\cal C}$. When
we act with ${\cal T}$, then
the combined power has to be the same modulo two, e.g.
${\cal C}^{mod_2 j}\leftrightarrow{\cal TC}^{mod_2(j-1)}$.
There is a twist. Namely, for
states with mixed even and odd excitation numbers like $|oeo\rangle$,
there is an additional sign, which shifts the relative power under ${\cal T}$,
so we have ${\cal C}^{mod_2 j} \leftrightarrow {\cal TC}^{mod_2 j}.$
So the subgroup ${\cal B}$ has two subsets even and odd under
${\cal TC}$-parity,
${\cal B}^e$ and ${\cal B}^o$. We call their members (i.e.~combinations
like ${\cal I}^i{\cal T}^t{\cal C}^c$) ${\cal Z}_j^e$ and  ${\cal Z}_j^o$.
At $r=4$ we have the pseudo-commutation relations\footnote{In full detail they read
\bea
 {\cal S}_1 {\cal C} =  {\cal I}  {\cal T}  {\cal S}_2 &&
 {\cal S}_1 {\cal C}^2 =  {\cal I}  {\cal T} {\cal C}{\cal S}_1 \quad\quad\quad
 {\cal S}_1 {\cal C}^3 = {\cal C}^3 {\cal S}_2\nonumber\\
 {\cal S}_1  {\cal T}   =  {\cal I} {\cal C}{\cal S}_2 \quad\quad\quad
 {\cal S}_1  {\cal T} {\cal C} =  {\cal I} {\cal C}^2 {\cal S}_1 &&
 {\cal S}_1  {\cal T} {\cal C}^2 =  {\cal T} {\cal C}^2 {\cal S}_2 \quad\quad\quad
 {\cal S}_1  {\cal T} {\cal C}^3 =  {\cal T} {\cal C}^3 {\cal S}_1\nonumber\\
 {\cal S}_1  {\cal I}   =  {\cal I}  {\cal S}_1 \quad\quad\quad
 {\cal S}_1  {\cal I} {\cal C} =  {\cal T}  {\cal S}_2 &&
 {\cal S}_1  {\cal I} {\cal C}^2 =  {\cal T} {\cal C}{\cal S}_1 \quad\quad\quad
 {\cal S}_1  {\cal I} {\cal C}^3 =  {\cal I} {\cal C}^3 {\cal S}_2\nonumber\\
 {\cal S}_1  {\cal I}  {\cal T}   = {\cal C}{\cal S}_2 \quad\quad\quad
 {\cal S}_1  {\cal I}  {\cal T} {\cal C} = {\cal C}^2 {\cal S}_1 &&
 {\cal S}_1  {\cal I}  {\cal T} {\cal C}^2 =  {\cal I}  {\cal T} {\cal C}^2 {\cal S}_2 \quad\quad\quad
 {\cal S}_1  {\cal I}  {\cal T} {\cal C}^3 =  {\cal I}  {\cal T} {\cal C}^3 {\cal S}_1\nonumber\\
 {\cal S}_2 {\cal C} = {\cal C}{\cal S}_1 &&
 {\cal S}_2 {\cal C}^2 =  {\cal I}  {\cal T} {\cal C}^3 {\cal S}_2 \quad\quad\quad
 {\cal S}_2 {\cal C}^3 =  {\cal I}  {\cal T}  {\cal S}_1\nonumber\\
 {\cal S}_2  {\cal T}   =  {\cal I} {\cal C}^3 {\cal S}_1\quad\quad\quad
 {\cal S}_2  {\cal T} {\cal C} =  {\cal T} {\cal C}{\cal S}_2 &&
 {\cal S}_2  {\cal T} {\cal C}^2 =  {\cal T} {\cal C}^2 {\cal S}_1\quad\quad\quad
 {\cal S}_2  {\cal T} {\cal C}^3 =  {\cal I} {\cal C}^2 {\cal S}_2\nonumber\\
 {\cal S}_2  {\cal I}   =  {\cal I}  {\cal S}_2\quad\quad\quad
 {\cal S}_2  {\cal I} {\cal C} =  {\cal I} {\cal C}{\cal S}_1 &&
 {\cal S}_2  {\cal I} {\cal C}^2 =  {\cal T} {\cal C}^3 {\cal S}_2\quad\quad\quad
 {\cal S}_2  {\cal I} {\cal C}^3 =  {\cal T}  {\cal S}_1 \nonumber\\
 {\cal S}_2  {\cal I}  {\cal T}   = {\cal C}^3 {\cal S}_1 \quad\quad\quad
 {\cal S}_2  {\cal I}  {\cal T} {\cal C} =  {\cal I}  {\cal T} {\cal C}{\cal S}_2 &&
 {\cal S}_2  {\cal I}  {\cal T} {\cal C}^2 =  {\cal I}  {\cal T} {\cal C}^2 {\cal S}_1\quad\quad\quad
 {\cal S}_2  {\cal I}  {\cal T} {\cal C}^3 = {\cal C}^2 {\cal S}_2.\nonumber
\eea}
\bea
  {\cal S}_i{\cal Z}_j^e&=&{\cal Z}_j^e{\cal S}_i,
  \quad\quad\quad\quad\mbox{(identity preserving)}\label{CR_IdPreserved}\\
            {\cal S}_i{\cal Z}_j^e&=&{\cal Z}_j^e{\cal S}_{mod'_2(i+1)}.
            \quad\mbox{(identity swapping)}
\eea
Note that
\[
  {\cal S}_i{\cal B}^e={\cal B}^e{\cal S}_i\quad \mbox{ and } \quad
  {\cal S}_i{\cal B}^o={\cal B}^o{\cal S}_{mod'_2(i+1)},
  \]
i.e.~while the individual operators do not pseudo-commute, they remain
in the same subset. At $r=4$ the following additional identities hold
\bea\label{r4Identities}
  {\cal S}_1{\cal S}_2&=&T{\cal Z}^o_k{\cal S}_1=T{\cal S}_2{\cal Z}^o_k,\\
  {\cal S}_2{\cal S}_1&=&T{\cal Z}^o_k{\cal S}_2=T{\cal S}_1{\cal Z}^o_k,
  \nonumber
\eea
for some $k$, so that ${\cal Z}^o_k\in{\cal B}^o$. In fact,
${\cal Z}^o_k={\cal TC}^2$. In other words, the
product of the two ${\cal S}$-operators can be reduced to the leading one,
but an odd ${\cal Z}$ operator appears as well as an additional ${\cal T}$
operator which switches the $T$ sign.
We are now ready to calculate the action of ${\cal S}_i$ on the totally
symmetrized state. There are two versions of that
state. If focusing on ${\cal S}$ symmetry properties, it is natural
to write the state in terms of {\em left} coset statelets ${\cal SB}|r\rangle$
\[
  {\cal G}|r=4\rangle = \left({\cal B}^e +{\cal B}^o\right)|r\rangle
  +S{\cal S}_1\left({\cal B}^e +{\cal B}^o\right)|r\rangle
  +S{\cal S}_2\left({\cal B}^e +{\cal B}^o\right)|r\rangle.
\]
On the other hand, we have to preserve cyclical properties due to the
structure of the Hamiltonian, in which case we should use {\em right}
coset statelets ${\cal BS}|r\rangle$\footnote{If this seems inconsistent,
  recall that the fully symmetrized
  state is symmetrized both in ${\cal B}$ and in ${\cal S}$. Hence, though parts
  of the state will appear to be symmetrized with respect to only one symmetry,
  the state as a whole has to be symmetric with respect to both
  operations. In general, this is only possible for certain combinations
  of quantum numbers. This is the reason why many sectors of the theory do
  not give rise to proper states.}
\[
  {\cal G}|r=4\rangle = \left({\cal B}^e +{\cal B}^o\right)|r\rangle
  +S\left({\cal B}^e +{\cal B}^o\right){\cal S}_1|r\rangle
  +S\left({\cal B}^e +{\cal B}^o\right){\cal S}_2|r\rangle.
\]
Both requirements lead to the same constraint on the quantum numbers.
In the latter case we have     
\bea
    {\cal S}_1{\cal G}|r=4; eee\rangle &=&
    {\cal S}_1\left({\cal B}^e +{\cal B}^o\right)|r\rangle
  +S{\cal S}_1\left({\cal B}^e +{\cal B}^o\right){\cal S}_1|r\rangle
  +S{\cal S}_1\left({\cal B}^e +{\cal B}^o\right){\cal S}_2|r\rangle\nonumber\\
    &=&
    \left({\cal B}^e{\cal S}_1 +{\cal B}^o{\cal S}_2\right)|r\rangle
  +S\left({\cal B}^e +{\cal B}^o{\cal S}_2{\cal S}_1\right)|r\rangle
  +S\left({\cal B}^e{\cal S}_1{\cal S}_2 +{\cal B}^o\right)|r\rangle\nonumber\\
    &=&
    S\left({\cal B}^e+{\cal B}^o\right)|r\rangle
  +\left({\cal B}^e +ST{\cal B}^o\right){\cal S}_1|r\rangle
  +\left(ST{\cal B}^e +{\cal B}^o\right){\cal S}_1|r\rangle\nonumber\\
  &=& S{\cal G}|r=4; eee\rangle \quad \mbox{(if $ST=1$),}
\eea
where we used Eq.~(\ref{r4Identities}) in the last step.
The calculation goes analogously for ${\cal S}_2$.
We thus proved that a symmetrized state is an eigenstate of ${\cal S}_1$
(and ${\cal S}_2$) in the $|eee; T\pm S\pm\rangle$ sector of the theory.
In the mixed sector, the ${\cal C}^j$ and ${\cal TC}^{j-1}$
terms get an extra sign for odd $j$ due to $(-1)^n=(-1)^l=-1$.
This effectively reverses $T$, therefore the viable sectors are
$|oeo; T\mp S\pm\rangle$. 

If we organize in terms of the left cosets we come to the same conclusion.
To wit 
\bea
    {\cal S}_1{\cal G}|r=4; eee\rangle &=&
    {\cal S}_1\left({\cal B}^e +{\cal B}^o\right)|r\rangle
  +S\left({\cal B}^e +{\cal B}^o\right)|r\rangle
  +S{\cal S}_1{\cal S}_2\left({\cal B}^e +{\cal B}^o\right)|r\rangle\nonumber\\
    &=&
    {\cal S}_1\left({\cal B}^e +{\cal B}^o\right)|r\rangle
  +S\left({\cal B}^e +{\cal B}^o\right)|r\rangle
  +ST{\cal S}_2{\cal Z}^o\left({\cal B}^e +{\cal B}^o\right)|r\rangle\nonumber\\
    &=&
    S\left[\left({\cal B}^e+{\cal B}^o\right)|r\rangle
  +S{\cal S}_1\left({\cal B}^e +{\cal B}^o\right)|r\rangle
  +T{\cal S}_2\left({\cal B}^e +{\cal B}^o\right)|r\rangle\right]\nonumber\\
  &=& S {\cal G}|r=4; eee\rangle \quad \mbox{(if $S=T$).}
\eea
The $I$ quantum number is necessarily the opposite of $T$. The reason
is that in the four-parton sector the pseudo-commutation relations
are such that ${\cal TI}$-parity $(-1)^{t+i}$ is conserved: commuting a
(primary) ${\cal S}$-operator with a ${\cal B}$ operator will yield
a different ${\cal B}$ operator with equal $t+i$ (modulo two). This
essentially links the identity and ${\cal T}$ statelets of ${\cal S}_1$ to the
${\cal IT}$ and ${\cal I}$ statelets of ${\cal S}_2$, respectively. 
Therefore $T$ and $I$ have to be different\footnote{In fact,
  %the ``left coset condition'' on $S$ and
  %the ``right coset condition'' on $I$
  this yields exactly two sectors, which makes
  sense in the even $r$ sectors, since here we have two excitation number
  sectors (even and mixed) for the required four (two massive {\em and} massless
  sectors. In the odd $r$ sectors there is only one condition due to the
  fact that there is no sign due to relative ${\cal C}$ operators; at
  odd $r$ all cyclic permutations enter with a positive sign.},
because otherwise the wavefunction
is identically zero.

The higher parton sectors, $r>4$, are harder to analyze because 
the pseudo-commutation relations
are more complicated. In particular, there is no identity
preserving relation akin to Eq.~(\ref{CR_IdPreserved}). Rather, we have
to use Eqs.~(\ref{pseudoCR}). Here ${\cal I}$ is decoupled, and
$I$-parity thus independently conserved.
So we have to come up with an ``operator calculus'' that allows us to evaluate
the $S$ sign of a combination of operators. As it turns out, $r=5$ is
not the most general case, but tractable and yielding clues
for the generic case $r>5$, so we discuss it here.

The exhaustively symmetrized state 
in terms of {\em left} coset elements can formally be written as
\[
  {\cal G}|r\rangle = {\cal B}|r\rangle
  %+S\sum_{i=1}^{r}{\cal S}_i{\cal B}|r\rangle
  +\sum_{j=1}^{N(r)}S_j{\cal S}_j{\cal B}|r\rangle
= \Big(1+\sum_{j=1}^{N(r)}S_j{\cal S}_j\Big){\cal B}|r\rangle,
\]
where the $r$ primary operators ${\cal S}_j$ carry an $S$ sign, so $S_j=S$
for $j\le r$. The
non-primary operators ${\cal S}_j$ carry an unknown sign $S_j=\pm S$. As
before, $N(r)=\frac{1}{2}(r-1)!-1=11,59,359,\ldots$.
The action of a primary operator ${\cal S}_k$ on an
exhaustively symmetrized state is then
\[
  {\cal S}_k{\cal G}|r\rangle = \Big( {\cal S}_k +S
  +\sum_{j=1 \atop j\neq k}^{N(r)}S_j{\cal S}_k{\cal S}_j\Big) {\cal B}|r\rangle
  = S \Big( 1 + S{\cal S}_k
  +S\sum_{j=1\atop j\neq k}^{N(r)}S_j{\cal S}_k{\cal S}_j\Big){\cal B}|r\rangle,
\]
where we have used $S_k=S$ for primary operators.
Due to the group axioms,
the bi-operators ${\cal S}_k{\cal S}_j$ can be expressed in terms of a
single ${\cal S}$ operator followed by operators of the ${\cal B}$ subgroup,
to wit
\[
{\cal S}_k{\cal S}_j={\cal S}_l{\cal I}^i{\cal T}^t{\cal C}^c,
\]
where the index $l$ and the powers $i,t,c$ can be looked up in a table like
Table \ref{GroupMultTable_r5}. But to reproduce the exhaustively symmetrized
state ${\cal G}|r\rangle$, we need to have
\beq\label{SignCondition}
SS_jI^iT^tC^c=S_l,
\eeq
since the ${\cal B}$ operators can be absorbed into ${\cal B}|r\rangle$,
whence only the quantum numbers $I,T,C$ re-emerge. This is the condition
we can use to identify the $S$ signs of the non-primary operators.

As an example, consider $r=5$, where we can formally set $C=1$, since cyclic
permutations at odd $r$ do not carry signs. Acting with ${\cal S}_1$, say,
on ${\cal G}|5\rangle$ yields the operator indices and powers displayed
in the ${\cal S}_1$ column of Table \ref{GroupMultTable_r5}. Since $I^iT^t=1$
for all of them, Eq.~(\ref{SignCondition}) simplifies to $SS_j=S_l$, so that
$S_6=S_7=S_8=S_{10}=S_{11}=S^2$ and $S_9=S$.
Note the interplay between ${\cal S}_9$ and
${\cal S}_{10}$: since ${\cal S}_1{\cal S}_9\propto{\cal S}_{10}$ and
${\cal S}_1{\cal S}_{10}\propto{\cal S}_{9}$, one operator gets an $S$ sign,
and the other does not. Which one it is has actually to be decided by
acting with another primary operator on the exhaustively symmetrized state
or by writing the state in terms of {\em right} coset statelets.

%%%%%%%%%%%%%%%%%%%%%%%%%% Group Multiplication TABLE r=5 %%%%%%%%%%%%%%%

\begin{table}
\begin{small}
\centerline{
\begin{tabular}{|c|ccccc|}\hline
  & ${\cal S}_1$& ${\cal S}_2$& ${\cal S}_3$& ${\cal S}_4$& ${\cal S}_5$\\\hline
${\cal S}_{1}$ & $ id   $ &  ${\cal S}_{11} {\cal I}  {\cal T}   $ &  ${\cal S}_{7}  $ &  ${\cal S}_{11}{\cal C}^3 $ &  ${\cal S}_{7} {\cal I}  {\cal T}  $ \\
${\cal S}_{2}$ & ${\cal S}_{6}  $ &  $ id   $ &  ${\cal S}_{10}  $ &  ${\cal S}_{6} {\cal I}  {\cal T} {\cal C}^3 $ &  ${\cal S}_{10} {\cal I}  {\cal T} {\cal C}$ \\
${\cal S}_{3}$ & ${\cal S}_{7}  $ &  ${\cal S}_{8}  $ &  $ id   $ &  ${\cal S}_{7} {\cal I}  {\cal T} {\cal C}^4 $ &  ${\cal S}_{8} {\cal I}  {\cal T} {\cal C}$ \\
${\cal S}_{4}$ & ${\cal S}_{11}{\cal C}^3 $ &  ${\cal S}_{6} {\cal I}  {\cal T} {\cal C}^3 $ &  ${\cal S}_{11} {\cal I}  {\cal T} {\cal C} $ &  $ id   $ &  ${\cal S}_{6}{\cal C}^4$ \\
${\cal S}_{5}$ & ${\cal S}_{8}{\cal C}^4 $ &  ${\cal S}_{10} {\cal I}  {\cal T} {\cal C} $ &  ${\cal S}_{8} {\cal I}  {\cal T} {\cal C} $ &  ${\cal S}_{10}{\cal C} $ &  $ id  $ \\
${\cal S}_{6}$ & ${\cal S}_{2}  $ &  ${\cal S}_{4} {\cal I}  {\cal T} {\cal C}^3 $ &  ${\cal S}_{9} {\cal I}  {\cal T} {\cal C}^2 $ &  ${\cal S}_{2} {\cal I}  {\cal T} {\cal C}^3 $ &  ${\cal S}_{4}{\cal C}$ \\
${\cal S}_{7}$ & ${\cal S}_{3}  $ &  ${\cal S}_{9}  $ &  ${\cal S}_{1}  $ &  ${\cal S}_{3} {\cal I}  {\cal T} {\cal C}^4 $ &  ${\cal S}_{1} {\cal I}  {\cal T}  $ \\
${\cal S}_{8}$ & ${\cal S}_{5}{\cal C} $ &  ${\cal S}_{3}  $ &  ${\cal S}_{5} {\cal I}  {\cal T} {\cal C} $ &  ${\cal S}_{9} {\cal I}  {\cal T} {\cal C}^4 $ &  ${\cal S}_{3} {\cal I}  {\cal T} {\cal C}$ \\
${\cal S}_{9}$ & ${\cal S}_{10} {\cal I}  {\cal T} {\cal C}^2 $ &  ${\cal S}_{7}  $ &  ${\cal S}_{6} {\cal I}  {\cal T} {\cal C}^2 $ &  ${\cal S}_{8} {\cal I}  {\cal T} {\cal C}^4 $ &  ${\cal S}_{11} $ \\
${\cal S}_{10}$ & ${\cal S}_{9} {\cal I}  {\cal T} {\cal C}^2 $ &  ${\cal S}_{5} {\cal I}  {\cal T} {\cal C} $ &  ${\cal S}_{2}  $ &  ${\cal S}_{5}{\cal C}^4 $ &  ${\cal S}_{2} {\cal I}  {\cal T} {\cal C}$ \\
${\cal S}_{11}$ & ${\cal S}_{4}{\cal C}^2 $ &  ${\cal S}_{1} {\cal I}  {\cal T}   $ &  ${\cal S}_{4} {\cal I}  {\cal T} {\cal C} $ &  ${\cal S}_{1}{\cal C}^2 $ &  ${\cal S}_{9} $ \\
\hline
\end{tabular}}
\end{small}
\caption{The ${\cal E}$ part of the group multiplication table in the
  five-parton sector organized as left coset elements.
  Since there is a small number of primary operators, the
  table has been transposed, in the sense that the column-heading operators
  act on the row-leading operators, e.g.~${\cal S}_1{\cal S}_2={\cal S}_6$.
  It is obvious that ${\cal E}$ is not a subgroup of ${\cal G}$;
  multiplying ${\cal S}$ operators generates ${\cal B}$ operators. 
\label{GroupMultTable_r5}}
\end{table}

%%%%%%%%%%%%%%%%%%%%%%%%%%%%%%%%%%%%%%%%%%%%%%%%%%%%%

The general algorithm is then as follows. Initially, we produce an
exhaustive list of operators by acting with primary operators on primary
operators until no new operators arise. This leads to a
group multiplication table, such as Table \ref{GroupMultTable_r5} for $r=5$
which can be used to write down consistent expressions
for primary operators acting on the generic state (with yet undetermined
$S$ signs) in terms of left or right coset statelets. The condition that
the emerging state has to be an eigenstate of the primary operators determines
the signs, and may lead to a further condition on $I$ as a function of $T$.
Using the sign condition Eq.~(\ref{SignCondition}) we
start with the $r$ known signs $S_j$ of the primary operators to get another
$r$ signs $S_l$, and iterate the process. The only time this goes bad
is when both $S_j$ and $S_l$ are unknown, as in the case of ${\cal S}_{9}$ and
${\cal S}_{10}$ at $r=5$. Then we go on to the next ${\cal S}_{k}$. Another
constraint arises from the different powers of $I$ and $T$ in
the group multiplication table. It will
lead to a condition which determines one of these quantum numbers in terms
of the other, e.g.~$I=T$ at $r=5$. At even $r$ there will be another condition
linking (the value of) $S$ and $T$, leading to two viable sectors per (even
or mixed) excitation number sector, which is what we need to produce
exactly four sectors of bound states at every $r$.

\section{Sectors, Signs, and Sines}
\label{AppxSectorSigns}

The assignment of symmetry sectors can be confusing due to several
signs and quantum numbers involved. Let's straighten things out by considering
the simplest case, $r=2$. Here, we have $\phi_2(x_1,x_2)\doteq|n\rangle$,
so we think of the {\em wavefunction} $\phi_2$ depending on two momentum
fractions $x_k$, as being represented by an abstract vector $|n\rangle$
labeled by {\em one} excitation number.
The orientation symmetry ${\cal T}$ of the Hamiltonian acts on {\em states}
only via their fermionic operators $b_{ij}(-x_k)$.
By flipping color indices, ${\cal T}$ effectively
reverses the order of the operators and thus the order of momentum fractions.
%$(x_1,x_2, \ldots, x_r)\rightarrow (x_r,x_{r-1}, \ldots, x_1)$.
To compensate, also the
wavefunction must be rewritten with the last momentum fraction now being
first. Wavefunctions might be even or odd under this reversal, and
the trace of operators might or might not acquire a sign under ${\cal T}$.
To assign a state to a {\em sector}, we have to take into account
both behaviors.
For example, a two-parton state looks like\footnote{The integral only
  runs to $\frac{1}{2}$ ($\frac{1}{r}$ in general) since operator products
  with flipped momentum fractions are redundant under the trace. The relative
  sign between the two Fock states
  $\Tr[b(-x)b(1-x)]|0\rangle=-\Tr[b(1-x)b(x)]|0\rangle$ will
  give rise to a wavefunction odd under argument reversal here,
  but not in general.} 
\[
|\Phi_2\rangle= \frac{1}{N_c}\int_0^{\frac{1}{2}} dx
\phi_2(x,1-x)\Tr[b(-x)b(1-x)]|0\rangle. 
\]
Since
\[
{\cal T}b_{ij}(-x)b_{ji}(1-x)=b_{ji}(-x)b_{ij}(1-x)=\Tr[b(-x)b(1-x)]
\]
is ${\cal T}$ even without the need to reverse the arguments,
this state belongs to the ${\cal T}$ even sector, even though the
{\em wavefunction} $\phi_2$ is odd under
reversal of its arguments\footnote{Of course, we can also flip the $b$
  operators {\em and} the wavefunction arguments, with the same result.},
cf.~Eq.~(\ref{Cyclicity}). Technically, the
wavefunction is represented as 
\bea
|\phi_2\rangle&=& |n\rangle + T{\cal T}|n\rangle + I({\cal I}|n\rangle +
T{\cal I}{\cal T}|n\rangle)\nonumber\\
&=& |n\rangle + T(-1)^n|-n\rangle + I|-n\rangle + IT(-1)^n|n\rangle\nonumber,
\eea
and also as\footnote{We are deliberately sloppy here with normalization,
which will be taken care of by the Hilbert space integral,
Eq.~(\ref{HSVolume}).}
\beq\label{TwoPartonState}
|\phi_2\rangle = |n\rangle + (-1)^{r+1}{\cal C}|n\rangle
= |n\rangle - (-1)^n|-n\rangle,
\eeq
where we recalled that ${\cal T}|n\rangle={\cal C}|n\rangle=(-1)^n|-n\rangle$.
So for even $n$ we need $T=I$
and the vanishing of the wavefunction at $x=0$ is guaranteed by the
latter equation which produces a sine
wavefunction with$I=-1$. Clearly, the even $n$ states constitute
the massive sector. If $n$ is odd, on the other hand,
we get cosines with $I=1$. In both cases $T=-1$ even though the states
are ${\cal T}$ even, ${\cal T}|\phi_2\rangle=|\phi_2\rangle$.
In sum, its quantum number $T$ does not directly give away the ${\cal T}$
sector of a state\footnote{But almost, since
  ${\cal T}b_1b_2\cdots b_r=(-1)^{(r- \mod_2 r)/2}b_rb_{r-1}\cdots b_1$, we have
  ${\cal TG}|r\rangle=T(-1)^{(r-\mod_2r)/2}{\cal G}|r\rangle$,
  so a state with $T=+1$
  is a ${\cal T}$ even state in the $4,5,8,9,12,13,\ldots$ parton sectors.},
but a negative(positive) $I$ quantum number always
results in a wavefunction made of sines(cosines).

\end{appendix}

\end{document}